\documentclass[journal=jpccck,manuscript=article, reprint]{achemso}

\usepackage{chemformula} 
\usepackage[T1]{fontenc} 
\usepackage{graphicx} 
\usepackage{color} 
\usepackage{ulem}
\usepackage{amsmath,amssymb}
\usepackage{longtable}
\usepackage{lineno}
\usepackage{hyperref}

\SectionNumbersOn

\title{Effects of the growth kinetics on solute diffusion in porous films}

\author{Gabriela B. Correa} 
\email{gabriela_barreto@id.uff.br}

\author{Renan A. L. Almeida} 

\author{F\'abio D. A. Aar\~ao Reis} \email{reis@if.uff.br}

\affiliation{Instituto de F\'{i}sica, Universidade Federal Fluminense,
Avenida Litor\^{a}nea s/n, 24210-340 Niter\'{o}i, RJ, Brazil}

\keywords{thin films \sep growth \sep porosity \sep diffusion coefficient \sep tortuosity}

\begin{document}

\begin{abstract}

For the development of porous materials with improved transport properties, a key ingredient
is to determine the relations between growth kinetics, structure, and transport
parameters.
Here we address these relations by studying solute diffusion through three-dimensional
porous films produced by deposition models with controlled thickness and porosity.
A competition between pore formation by lateral particle aggregation
and surface relaxation that favors compaction is simulated by
the lattice models of ballistic deposition and of random deposition with surface relaxation,
respectively, with relative rates proportional to $p$ and $1-p$.
Effective diffusion coefficients are determined in steady state simulations with
a solute source at the basis and a drain at the top outer surface of the films.
For a given film thickness, the increase of the relative rate 
of lateral aggregation leads to the increase of the effective porosity and of the diffusivity,
while the tortuosity decreases.
With constant growth conditions, the increase of the film thickness always leads to the increase
of the effective porosity, but a nontrivial behavior of the diffusivity is observed.
For deposition with $p\leq 0.7$, in which the porosity is below $0.6$,
the diffusion coefficient is larger in thicker films;
the decrease of the tortuosity with the thickness quantitatively confirms
that the growth continuously improves the pore structure for the diffusion.
Microscopically, this result is associated with narrower distributions of the local solute current
at higher points of the deposits.
For deposition with $p\geq 0.9$, in which the films are in the narrow porosity range
$\sim 0.65-0.7$, the tortuosity is between $1.3$ and $2$, increases with the thickness, and has maximal changes near $25\%$.
Pairs of values of porosity and tortuosity obtained in some porous electrodes 
are close to the pairs obtained here in the thickest films,
which suggests that our results may be applied to deposition of materials of
technological interest.
Noteworthy, the increase of the film thickness is generally favorable
for diffusion in their pores, and the exceptions have small losses in tortuosity.

\end{abstract}

\maketitle

\section{Introduction}
\label{intro}

The development of porous materials with improved transport properties is interesting for several
technological applications,
such as electrodes in energy storage devices \citep{liJOM2017,wangACSAEM2020},
membranes for ion or molecule separation \citep{koros2017,linCCRev2019},
and catalysis \citep{holek2019,zhengChemA2018}.
For instance, the design of high capacity Li-ion batteries
partly relies on the production of porous electrodes that are mechanically stable and
offer low resistance to electron and ion transport \citep{heubner2020}.
This may be achieved with better battery designs or with the development of novel materials.
In all cases, it is important to understand the relation between the conditions in which the
materials are formed, the geometry of their porous systems, and transport parameters such
as conductivities or diffusion coefficients.
Since electrodes are frequently produced in the form of thin films, it is reasonable to
search for those relations in this type of structure.

Several works have already addressed the relations between structural and
transport properties in porous media,
with interest not only on manufactured materials but also in natural ones such as rocks and soils
\citep{dullien,Adler}.
In most cases, the porous media are modeled with preset distributions of solid and pore space,
such as regular or random sphere packings \citep{guyon1987,kim1991,coelho1997},
fiber networks \citep{tomadakis1993},
tubes with cage-throat geometry \citep{holek2019,sund2017}, etc,
or use reconstructed images of the materials of interest \citep{muter2015,tallarek2019}.
Recent works modeled porous electrodes and performed
transport simulations using several methods
\citep{jiangApEn2017,torayev2018,elabyouki2019,tanPCCP2019}.
However, they do not account for the dynamical processes for porous media formation.

Here we advance in the determination of structure-transport relationships by
studying models that connect them with the conditions of film growth.
We study diffusional transport of a solute through films with broad ranges of
porosity and thickness, which are formed by sequential aggregation of particles that
may follow mechanisms of pore formation or of surface relaxation.
Tuning the relative rate of these mechanisms changes the growth kinetics and,
consequently, changes bulk and surface properties.
Besides analyzing the effects of changing growth conditions, we also investigate the effects of
film thickness and surface roughness on quantities such as porosity and tortuosity.
Among the results, we highlight the observation that the increase of the thickness is
beneficial for the diffusivity in all films with porosity below $60\%$,
with tortuosity reductions reaching factors close to $5$ when the porosity is near $20\%$.
The relations between porosity and tortuosity obtained here closely match the relations obtained
in some porous electrodes, which suggests that our results may be relevant for their production.

The competitive growth kinetics is simulated in three dimensions (3D) with two widely studied models:
ballistic deposition (BD) \citep{barabasi,vold},
in which pores are formed by lateral aggregation of particles,
and random deposition with surface relaxation (RDSR) \citep{family},
which favors the formation of compact structures.
The competitive BD-RDSR model can describe a variety of conditions for formation of porous films,
with the constraints of an at least partially
collimated particle flux and of negligible subsurface relaxation.
These mechanisms differ, for instance, from those of evaporation-induced 
nanoparticle assembly, which were recently modeled in two dimensions (2D) \citep{liuPCCP2017},
and from deposition with flux distributed in several directions \citep{lehnen2010}.

The rest of this paper is organized as follows.
In Sec. \ref{basics}, we present the deposition model, the model of diffusive transport,
the quantities of interest, and a brief review of previous results on the BD-RDSR model.
In Sec. \ref{results}, we show numerical results for the structural properties of the deposits,
the effective diffusion coefficients in their pores, and the distributions of local solute currents
in their cross sections.
In Sec. \ref{discussion}, we analyze the relations between the growth conditions, the film structure,
and the transport parameters.
In Sec. \ref{conclusion}, we summarize our results and present our conclusions.
A list of symbols used in this paper is presented in Table \ref{NomenclatureTable}.

\section{Model and methods}
\label{basics}

\subsection{Film deposition}
\label{modelfilm}

The deposit has simple cubic lattice structure with sites of edge length $a$.
Deposition occurs on a flat substrate located at $z=0$, with lateral size $L$,
and periodic boundary conditions in $x$ and $y$ directions.
Each deposited particle occupies one lattice site.
Depending on the application of the model, a particle may represent a molecule, an aggregate,
or a nanoparticle.
The set of particles with the same $\left( x,y\right)$ positions is termed a column of the deposit.
The height $h\left( x,y\right)$ is the maximal height $z$ of a particle in a column and
the set $\{h\left( x,y\right)\}$ defines the top outer surface.

\begin{figure}[!ht]
\center
\includegraphics[clip,width=0.4\textwidth,angle=0]{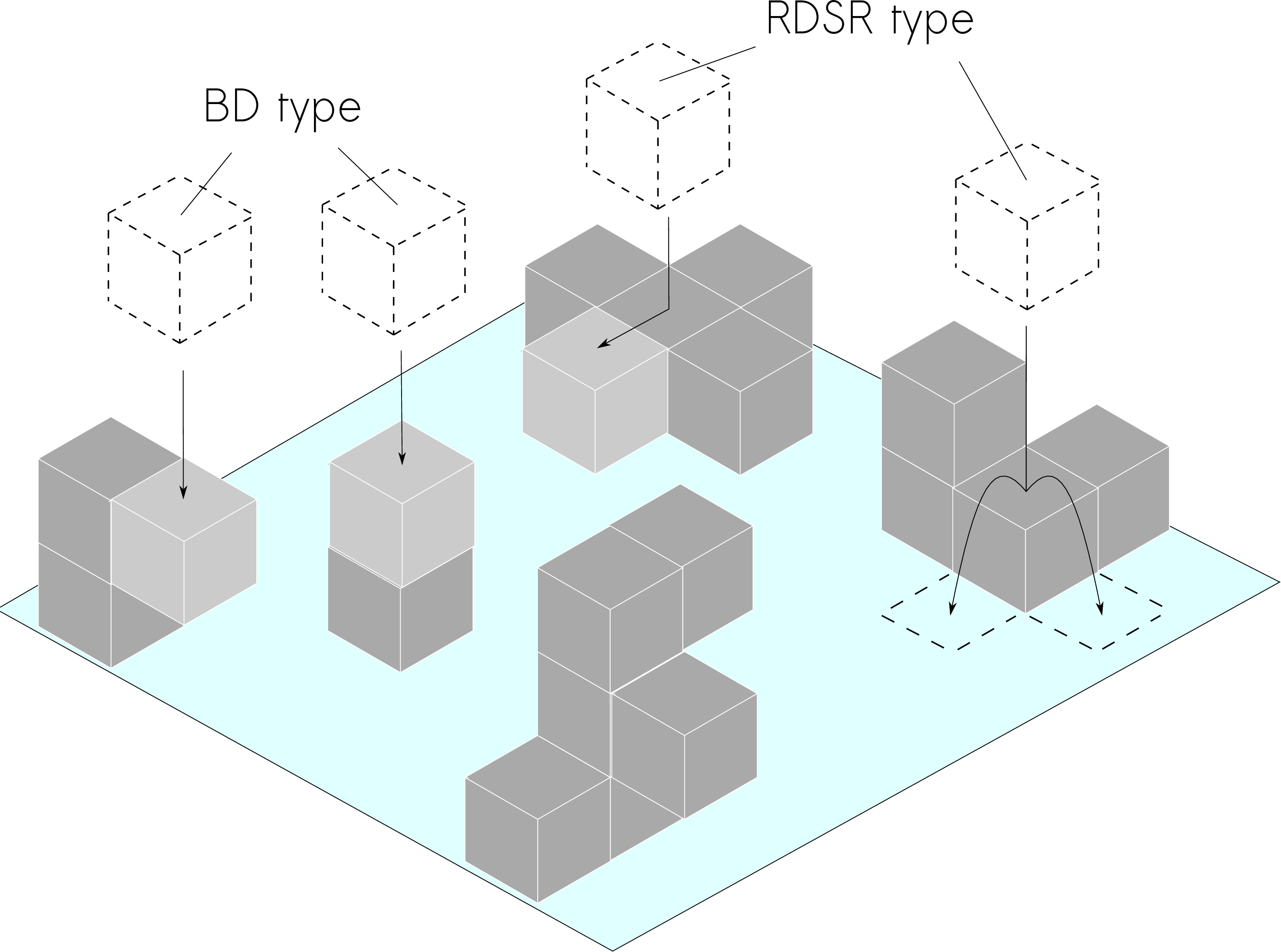}
\caption{
Scheme of the aggregation mechanisms of BD and RDSR.
Dashed cubes are incoming particles, dark gray cubes are aggregated particles,
and light gray cubes show the aggregation position of some incoming particles.
If more than one aggregation position can be chosen, they are indicated by the split arrows.
}
\label{figmodel}
\end{figure}

Incident particles are sequentially released in randomly chosen columns $(x,y)$, with $z$
larger than all $h\left( x,y\right)$, and follow a trajectory in the $-z$ direction.
The rules for their aggregation are illustrated in Fig. \ref{figmodel}.
With probability $p$, the particle aggregates according to the BD rule, namely
at its first contact with a previously deposited nearest neighbor (NN) \citep{vold,barabasi}.
The aggregation in a lateral contact (e.g. at the left in Fig. \ref{figmodel})
either creates a pore or expands an existing pore.
With probability $1-p$, the incident particle follows the RDSR rule:
it attaches at the top of the column of incidence if no NN column has a smaller
height, otherwise it moves to the top of the NN column with the minimal height;
in the case of a draw, one of the columns with minimal height is randomly chosen \citep{family}.
The aggregation follows the same rules when the particle is adsorbed on the substrate and
in the deposit.

\begin{figure}[!ht]
\center
\includegraphics[clip, width=0.25 \textwidth, angle=0]{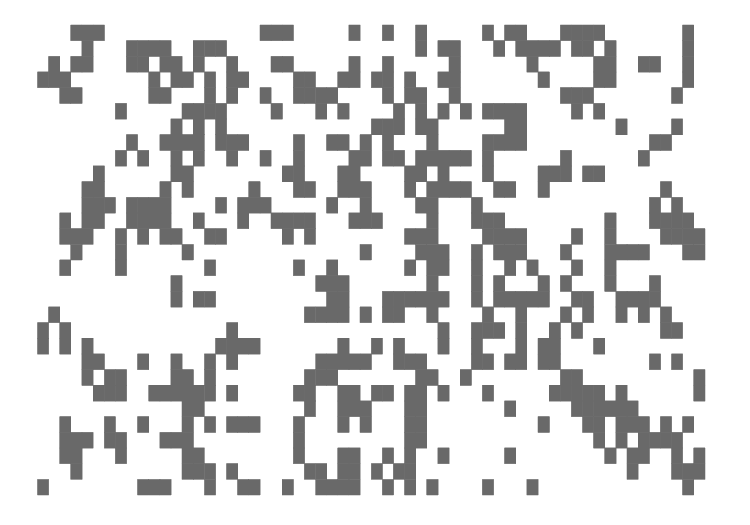} 
\includegraphics[clip, width=0.25 \textwidth, angle=0]{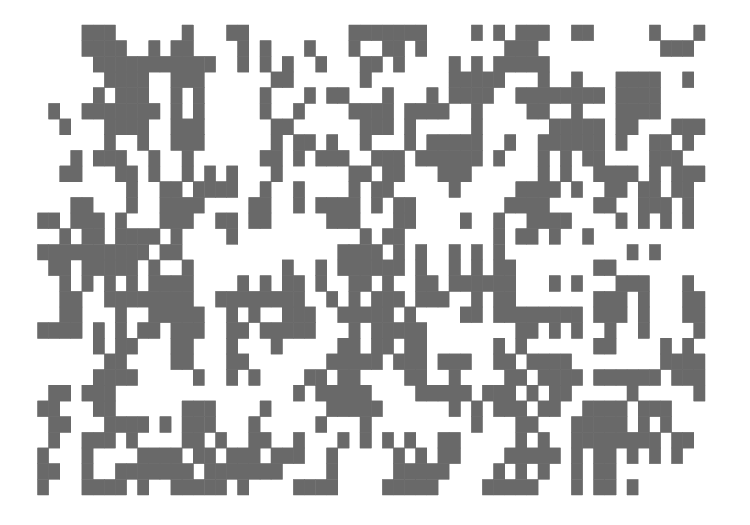} 
\includegraphics[clip, width=0.25 \textwidth, angle=0]{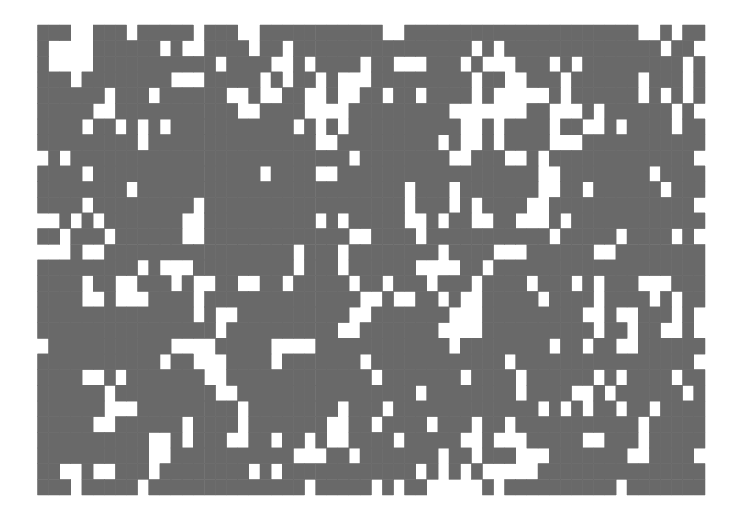} 

\vspace{0.1cm}

\includegraphics[clip, width=0.2 \textwidth, angle=0]{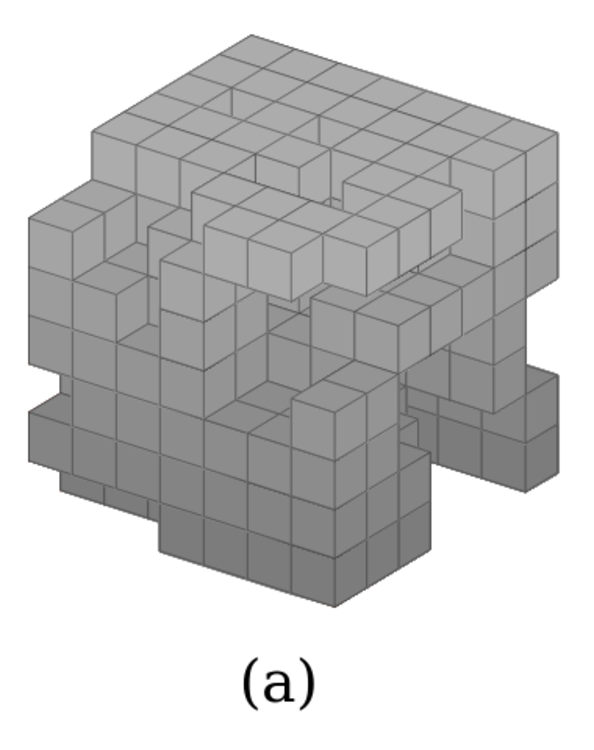} \qquad
\includegraphics[clip, width=0.2 \textwidth, angle=0]{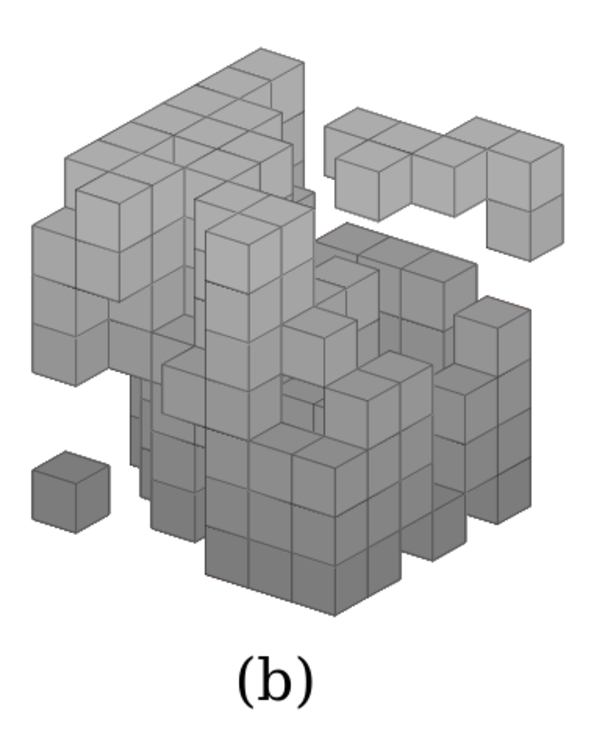} \qquad
\includegraphics[clip, width=0.2 \textwidth, angle=0]{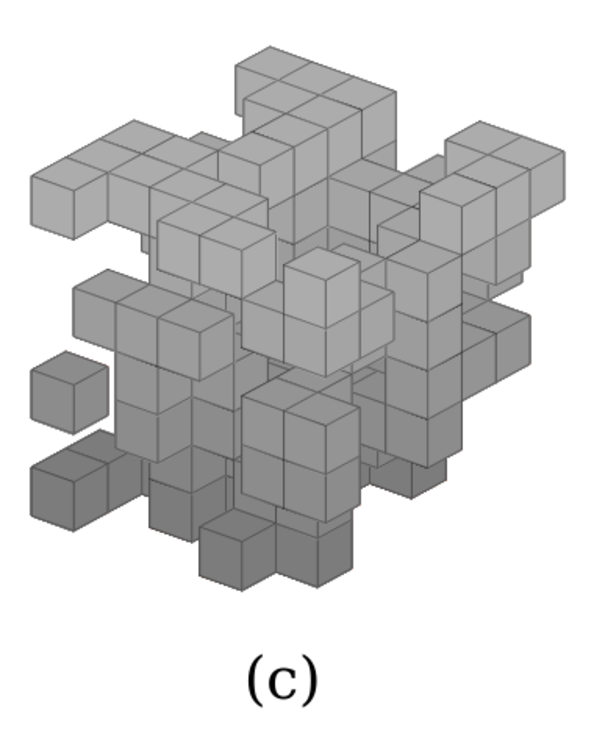} \qquad
\caption{
Horizontal cross sections and small parts of the 3D deposits grown with
(a) $p = 0.4$, (b) $0.7$, and (c) $1$.
}
\label{examples}
\end{figure}

Figs. \ref{examples}(a)-(c) show horizontal cross sections and small 3D parts
of the deposits produced with some values of $p$.
Close inspection indicates that the porosity and the connectivity increase with $p$.
Details of the simulations of film growth and calculation of structural properties
are in Sec. SI.I of the Supporting Information.
  
The BD-RDSR model is not intended to describe a particular deposition process, but it is
designed to be an approximation of thin film deposition processes with competitive
adsorption mechanisms.
This competition is simplified by aggregation at first contact, with relative rate $p$,
and surface relaxation, with relative rate $1-p$.
For instance, in some sputtering processes, the energy distribution of ions,
molecules, or clusters may be broad, so that the aggregation at the first contact is more likely
for the species with low energy, while those with high energy tend to form locally
compact configurations.
In this case, the energy distribution depends on the interactions of ions and the target source. 
Similar interpretation is possible in electrospray deposited films if there are differences in the
nanoparticle aggregation rates due to charge distributions.
In cases of deposition of a mixture with a binder, the concentration of this binder may control
the rate of pore formation; this is done in the production of battery electrodes.
Thus, it is frequently possible to infer which changes in the growth conditions will increase
or decrease the effective value of $p$, although it is difficult to estimate that
value directly from the growth parameters (this estimation may be done by comparing
the structural properties that are calculated here).
Another important feature of the BD-RDSR model is to describe short range surface relaxation
processes, i.e. relaxation only to NNs.
This is expected to be a reasonable approximation for low temperatures and high deposition fluxes,
in which adsorbed particles or aggregates have small surface mobility.

When $N_d$ is the number of deposited particles, the
dimensionless growth time is defined as the number of particles per column,
$t_g=N_d/{\left( L/a\right)}^2$.

\subsection{Structural properties}
\label{methodmorphology}

The dimensionless thickness $H$ is the average of the dimensionless heights $\{h/a\}$ of the top outer
surface:
\begin{equation}
    H =
    \left\langle \frac{1}{{\left( L/a\right)}^2}\sum_{x,y}{\frac{h\left( x,y\right)}{a}} \right\rangle ,
    \label{defH}
    \end{equation}
where the angular brackets denote an average
over different deposits with the same growth time $t_g$.
Fluctuations of the outer surface profile are quantified by the dimensionless roughness
    \begin{equation}
    W = \Bigg \langle \sqrt{\frac{1}{\left( L/a\right)^2}
    \sum_{x,y}{\left[\frac{h\left( x,y\right)}{a}-H\right]^2}} \Bigg \rangle.
    \label{defW}
    \end{equation}

The deposit is limited by five flat
surfaces, namely the substrate and the four lateral sides, and the top outer surface,
so its total volume is $V_T=L^2 Ha$.
The total porosity $\epsilon_T$ is defined as the pore volume fraction:
\begin{equation}
\epsilon_T = 1-\frac{N_d a^3}{V_T} = 1-\frac{t_g}{H} = 1-\frac{1}{v_a} ,
\label{defporT}
\end{equation}
where $v_a=H/t_g$ is a dimensionless growth velocity averaged over the total growth time.

\begin{figure}[!ht]
\center
\includegraphics[clip,width=0.4\textwidth,angle=0]{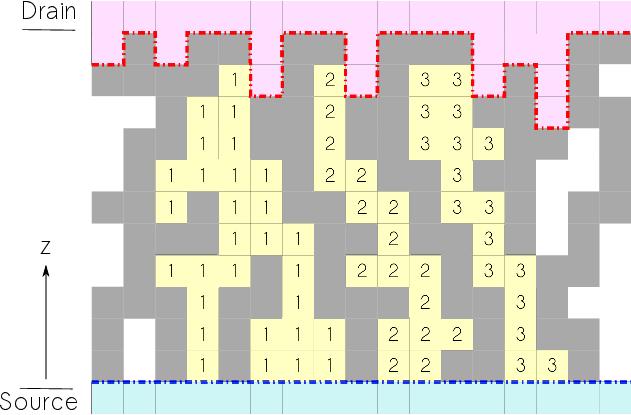}
\caption{
Sketch of a two-dimensional film with solid particles (gray),
a connected pore system (yellow), and unconnected pores (white).
The black dash-dotted line indicates the frontier of the source (blue)
and the red dash-dotted line indicates the frontier of the drain (rose).
Connected porous domains are labeled with numbers.
}
\label{effective}
\end{figure}

The effective porosity or connected porosity $\epsilon_E$ is the volume fraction that can be
filled by a fluid or a solute transported through the porous medium.
Here, the substrate is the frontier where the solute flows in, with a source below it,
and the top outer surface is the frontier where the solute flows out, with the drain
specified as $\{z(x,y)\textrm{  }|\textrm{  }z > h(x,y),\textrm{  } \forall\textrm{  }(x,y)\}$.
Fig. \ref{effective} shows a 2D porous deposit and indicates the two frontiers,
the source, and the drain (this facilitates visualization in comparison with 3D images).
We also assume that two pores are connected if they are NN and define 
the connected pore system as the set of internal pores that are connected to both frontiers
by some sequence of NN internal pores; this system is also indicated in Fig. \ref{effective}.
Denoting its volume as $V_c$, the effective porosity is
\begin{equation}
\epsilon_E \equiv \frac{V_c}{V_T} .
\label{defporE}
\end{equation}
Isolated pores are those that do not belong to the connected pore system.

A connected domain of the pore system is defined by:
$(i)$ it is a pore set connected to both frontiers;
$(ii)$ all pairs of sites in the domain are also connected by a sequence of NN pores of that domain.
Fig. \ref{effective} shows three connected domains in a 2D deposit.
The average number of connected domains is used here to quantify the film structure.
Their enumeration and the calculation of $\epsilon_E$ were performed via 
a 3D version of the Hoshen-Kopelman algorithm \citep{hoshen,stauffer}.

\subsection{Transport simulations}
\label{methodtransport}

We simulate solute transport across the deposits after the growth has stopped,
using an extension of an infiltration model \citep{reis2016}.
The concentration $C$ in a pore site or in the source is discretized, with possible
values $0$ (empty) and $1$ (occupied).
For applications in which the pore volume may contain more than one molecule,
the local solute concentration has to be determined by averaging the occupation number
over a larger volume.

A constant value $C=1$ is set at the source and $C=0$ is set at the drain.
The molecules in the pore system and in the source execute random walks to NN
sites with a rate $\nu$, which is defined as the number of hop attempts per unit time.
Hop attempts to occupied pores or to solid sites are rejected and the molecule remains at
its current position.
When a molecule leaves the source, another molecule is immediatly inserted at its previous position,
and when a molecule crosses the top frontier, it is removed from the system;
this maintains the concentrations in the source and in the drain.
In the hydrodynamic limit, the average concentration obeys the diffusion equation with
the same boundary conditions \citep{reisvoller2019}.

\begin{figure}[!ht]
\center

 \includegraphics[clip,width=0.3\textwidth,angle=0]{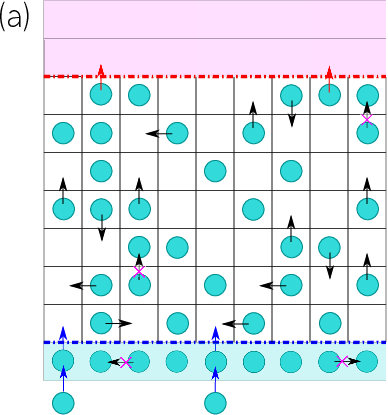} \quad
 \includegraphics[clip,width=0.3\textwidth,angle=0]{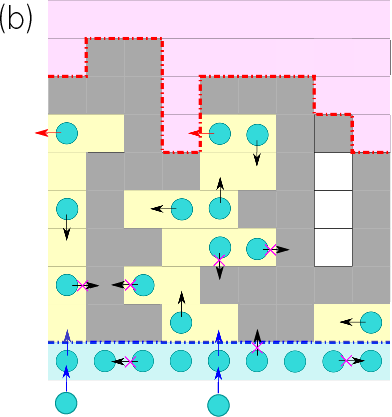}

\caption{
Two-dimensional illustration of the transport of solute molecules (circles) (a) in
a free medium and (b) in the pores of a deposit, using the same color scheme of
Fig. \ref{effective}.
Periodic conditions in the lateral boundaries are considered.
The arrows indicate hop attempts of some molecules and the rejected attempts
have arrows overlayed with a cross.
Particles that reach the drain are removed from the system (red arrows)
and particles that hop from the source to a pore are immediately replaced (blue arrows).}
\label{figtransport}
\end{figure}

Figs. \ref{figtransport}(a) and \ref{figtransport}(b) show 2D schemes of the model in
a free medium and in a porous deposit, respectively.

The currents $J_{in}$ and $J_{out}$ are defined as the average numbers of
molecules that cross the frontiers per unit time of infiltration (this time does not include
the growth time $t_g$).
After a transient, solute transport reaches a steady state, in which these currents
fluctuate around a constant value.
The stationary current $J_s$ is calculated by averaging them in time and over
different deposits.
The effective diffusion coefficient $D_E$ of the solute is expressed in terms of the coefficient
$D_0$ in the free medium as
\begin{equation}
D_r=\frac{D_E}{D_0} = 6\left(H+1\right) \frac{J_s a^2}{\nu} ,
\label{defDr}
\end{equation}
where $H+1$ stands for the average distance between the frontiers (see Fig. \ref{effective})
and $D_r$ is a normalized or relative effective diffusion coefficient.
Details on the method are in Sec. SI.II of the Supporting Information.

We also calculated the time-averaged solute concentrations
$\overline{C}\left( x,y,z\right)$ in the steady states of some deposits
and use them to determine the average solute current $\vec{j}_s$ as
\begin{equation}
\vec{j}_s = -a^2 \nu\vec{\nabla}{\left(\frac{\overline{C}}{a^3}\right)}
\label{defv}
\end{equation}
(note that $C$ was defined as a dimensionless concentration).
A dimensionless local current is defined as $\vec{j}=\left(a^2/\nu\right)\vec{j}_s$;
distributions of its absolute value, $j(x,y,z)$, and of its $z$ component, $j_z(x,y,z)$,
are used to analyze local transport properties.

\subsection{Previous results on the BD-RDSR model}
\label{previousworks}

As a film grows in a wide substrate, the roughness is frequently observed to increase
as a power law in time:
$W \sim t_g^\beta$, where $\beta$ is the growth exponent \citep{barabasi,krug}.
Surface fluctuations in the BD model are described in the hydrodynamic limit by the 
Kardar-Parisi-Zhang (KPZ) equation \citep{kpz}.
All 3D models in the KPZ class have $\beta \approx 0.243$ \citep{kelling2016}.
However, in simulations of BD and in other models with lateral particle aggregation,
the average slopes of $\log{W}\times \log{t_g}$ plots are usually smaller than $\beta$
for $t_g\lesssim {10}^5$.
This is ascribed to an intrinsic roughness $\xi_I$ that affects the roughness scaling as 
$W^2 \simeq \xi_I + At_g^{2\beta}$ ($A$ constant) \citep{kertesz,alvesPRE2014}.

The instantaneous growth velocity $v_g={\text d}H/{\text d}t_g$ in KPZ models varies in time as
\citep{krug1990}
\begin{equation}
v_g=v_\infty - \Lambda {t_g}^{-\alpha_\bot} ,
\label{vscaling}
\end{equation}
where $\alpha_\bot=1-\beta\approx 0.76$ in 3D, and $v_\infty$ and $\Lambda$ are constants.
In porous film growth, $\Lambda>0$, so the growth velocity slowly increases in time.

Surface fluctuations of the RDSR model are described by the Edwards-Wilkinson (EW)
equation \citep{ew,barabasi}, which 
implies that the roughness scales as $W^2\sim \ln{t_g}$.
No previous work on RDSR has shown large corrections to this scaling relation.

The BD-RDSR model was proposed to study a roughening transition between the EW
and the KPZ classes \citep{pellegrini}.
Nonperturbative renormalization group calculations discards such transition in
3D \citep{canetPRL2010}, which implies that the model is in the KPZ class for all $p > 0$.
However, deviations may be observed in simulations
due to e.g. the intrinsic roughness.
In BD-RDSR in 2D, long crossovers from EW to KPZ scaling were already shown
for small $p$ \citep{chamereis2002,muraca2004,silveira2012}.

Bulk properties of BD-RDSR deposits in 2D were formerly investigated \citep{kriston2016},
but no connected pore system emerges in this case;
inspection of images of ballistic deposits in 2D actually show that $\epsilon_E=0$
\citep{barabasi,katzav2006}.
However, BD-RDSR deposits in 3D show a transition in the percolation class \citep{stauffer}
at $p_c \approx 0.35$, i.e. there is a connected pore system for $p > p_c$  and there is
no such system for $p < p_c$ \citep{yuamar2002}.
This motivates the present study of films grown above $p_c$ as controlled templates
for porous media with $\epsilon_E>0$. 
Diffusion of adsorbed molecules was already studied in the pore walls of
3D films grown by BD ($p=1$) \protect\cite{reiscaprio2014}, 
but the study was constrained to transient regimes.
To our knowledge, no previous work has studied the coupling of
structural and transport properties in BD-RDSR films.

\section{Results}
\label{results}

\subsection{Film thickness and surface roughness}
\label{surface}

The instantaneous velocity has slow variation in time, as predicted in Eq. (\ref{vscaling}).
Thus, at long times, the average thickness $H$ has an approximately linear increase with $t_g$.
Details are shown in Sec. SI.III of the Supporting Information.

\begin{figure}[!ht]
\center
\includegraphics[clip,width=0.4\textwidth,angle=0]{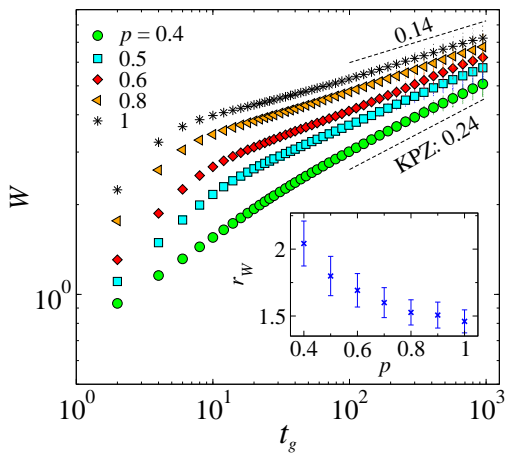}
\caption{
Roughness as function of the growth time for several values of $p$.
Dashed lines have the indicated slopes.
The inset shows the ratio $r_W$ as function of $p$.
The uncertainties are smaller than the size of the symbols in all plots.
}
\label{roughness}
\end{figure}

Fig. \ref{roughness} shows the time evolution of the roughness for several $p$.
At a fixed growth time $t_g$, the roughness increases with $p$,
which is expected because BD produces rougher deposits than RDSR.
The same is observed for a fixed thickness $H$.
The slope of the plots for $p\approx 1$ are near $0.14$, which differs from the KPZ value;
this can be related to the intrinsic width correction discussed in Sec. \ref{previousworks}.
However, as the BD aggregation is less frequent ($p \approx 0.4$), the average slope
becomes closer to the KPZ value $\beta\approx 0.24$.

To relate structural film properties, we analyze the variation of the roughness with the thickness.
For constant growth conditions, i.e. constant $p$, we calculate the ratio $r_W$ between the roughness
of relatively thick films, with $H = 10^3$, and of relatively thin films, with $H = 50$:
$r_W = W\left( H={10}^3\right) / W\left( H=50\right)$.
The inset of Fig. \ref{roughness} shows $r_W$ as a function of $p$.
The films with smoother surfaces, which are produced with $p\approx 0.4$,
have a faster increase in the roughness with the thickness (i.e. larger $r_W$)
in comparison with the rougher films produced with $p\approx 1$.

\subsection{Porosity and connectivity}
\label{porosity}

We begin with a brief review of the results of BD-RDSR in 2D \citep{kriston2016}.
For a fixed growth time or fixed thickness, the total porosity $\epsilon_T$ increases with $p$,
which is expected because lateral aggregation facilitates the formation of pores.
For constant $p$, Eqs. (\ref{defporT}) and the KPZ relation (\ref{vscaling}) for the growth velocity
explain why $\epsilon_T$ slowly increases with the thickness (the evolution of the 
average velocity $v_a$ is similar to that of the instantaneous velocity $v_g$).
However, all deposits in 2D have effective porosity $\epsilon_E=0$.
Thus, $\epsilon_E$ cannot be predicted from known relations of KPZ scaling
and have to be determined numerically.

\begin{figure}[!ht]
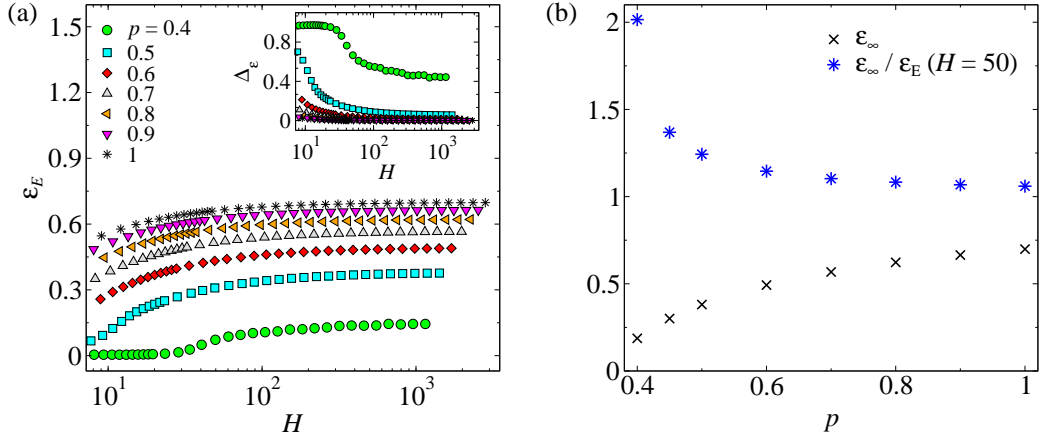

\center
\includegraphics[clip,width=0.4\textwidth,angle=0]{figporosity_a.eps} \quad
\includegraphics[clip,width=0.4\textwidth,angle=0]{figporosity_b.eps}
\caption{
(a) Effective porosity as function of the film thickness for several values of $p$.
The inset shows the relative difference between total and effective porosities.
(b) Effective porosity of very thick ($H \gtrsim 10^4$) films (crosses) and the ratio
$r_\epsilon$ (asterisks) as function of $p$.
The uncertainties are smaller than the sizes of the symbols in all plots.
}
\label{figporosity}
\end{figure}

In 3D, for constant $p$, we also observe the expected increase of $\epsilon_T$
with the thickness.
Fig. \ref{figporosity}(a) shows the same evolution for the numerically calculated
estimates of $\epsilon_E$ in cases where a connected pore system is formed, i.e. $p>p_c\approx0.35$;
for $p<p_c$, $\epsilon_E=0$ \citep{yuamar2002}.

The inset of Fig. \ref{figporosity}(a) shows the relative difference between the total and the
effective porosities,
$\Delta_\epsilon = \left(\epsilon_T - \epsilon_E\right) /\epsilon_T$,
which measures the fraction of the pore volume that cannot participate in the transport.
For $p\geq 0.7$ and $H>20$, this difference is small ($\lesssim 2\%$),
which means that isolated pores are seldom formed.
In the less porous films ($p=0.4-0.6$), non-negligible values of $\Delta_\epsilon$ persist
in large thicknesses.

$\epsilon_E$ has a nontrivial evolution for small thicknesses; see Fig. \ref{figporosity}(a).
After these transients, the slow variation allows us to extrapolate $\epsilon_E$
to the limit $H\to\infty$; see details in Sec. SI.IV of the Supporting Information.
For each $p$, this extrapolation gives an effective porosity $\epsilon_\infty$, which is representative
of films with $H \gtrsim {10}^4$ with an accuracy $\approx 1\%$; this is hereafter termed thick film limit.
Fig. \ref{figporosity}(b) shows that $\epsilon_\infty$ also increases with $p$,
extending the previous observations to much thicker films.

We also calculated the ratio
$r_\epsilon=\epsilon_\infty/\epsilon_E\left(H=50\right)$,
which quantifies the relative increase
in the porosity from thin film conditions ($H=50$) to the thick film limit.
This ratio is also shown in Fig. \ref{figporosity}(b).
The films with larger porosities are those that have
slower variations in the porosity with the thickness, i.e. have $r_\epsilon\approx 1$.
Instead, the films with relatively smaller porosity
show faster variations with the thickness; for instance, the porosity of films grown with $p=0.4$
increase more than $100\%$ from $H=50$ to the thick film limit.    

The evolution of the numbers of connected domains is shown in Sec. SI.V of the Supporting Information.
For $p\geq 0.5$, a single connected domain is found in thickness $H\approx 20$ or larger;
for $p=0.4$, it occurs in $H\geq 60$.
Thus, in all samples, there is a progressive merging of multiple percolating domains
into a single, large percolating pore cluster.

\subsection{Diffusion coefficients}
\label{diffusion}

\begin{figure}[!ht]
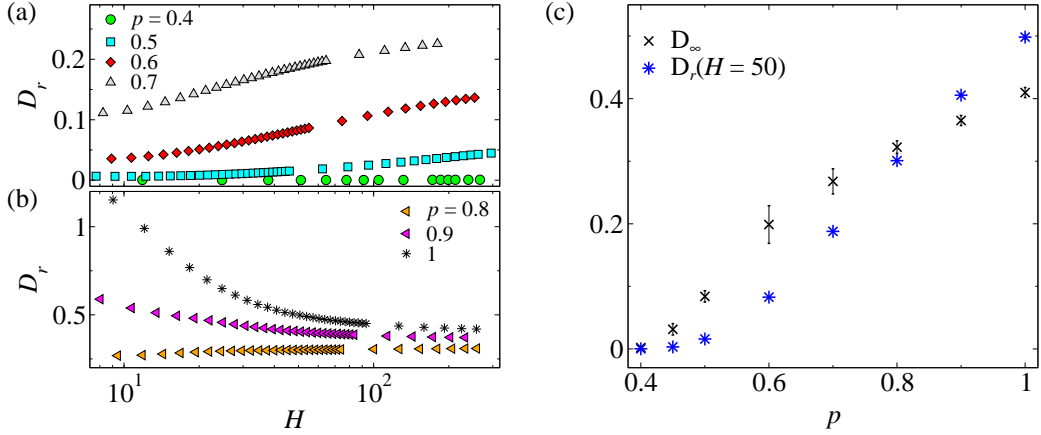

\center
\includegraphics[clip,width=0.4\textwidth,angle=0]{coef_ab.eps} \quad
\includegraphics[clip,width=0.4\textwidth,angle=0]{coef_c.eps}
\caption{ 
(a),(b): Normalized diffusion coefficient as function of film thickness for
small and large values of $p$, respectively.
(c) Normalized diffusion coefficients of very thick films, $D_{\infty}$,
and of thin films with $H=50$, $D_r\left( H=50\right)$, as a function of $p$.
Error bars of $D_{\infty}$ are shown, but uncertainties in the other
estimates are smaller than the sizes of the symbols.
}
\label{coef}
\end{figure}

Fig. \ref{coef}(a) shows the normalized diffusion coefficient
$D_r$ as a function of the film thickness for $p\leq 0.7$; 
Fig. \ref{coef}(b) shows the same quantities for $p>0.7$.
In contrast with the structural quantities, the evolution of $D_r$ shows
different trends: for $p<0.8$, $D_r$ increases with the thickness (the same as $\epsilon_E$);
for $p>0.8$, $D_r$ decreases with the thickness (the opposite of $\epsilon_E$).

The variations of $D_r$ also suggest to compare results in thin and thick films.
The effective diffusion coefficient in the thick film limit, $D_\infty$, is estimated 
using the same method adopted to calculate $\epsilon_\infty$;
see Sec. SI.IV of the Supporting Information.
$D_\infty$ is also a good approximation for the normalized diffusion coefficient in all films
with $H \gtrsim {10}^4$.
Fig. \ref{coef}(c) shows $D_\infty$ and the normalized diffusion coefficient in relatively thin films,
$D_r\left(H=50\right)$, as a function of $p$.
It confirms the nontrivial behavior for $p\geq0.9$, in which the diffusivity decreases
with the thickness; these films have porosity between $0.6$ and $0.7$ for all thicknesses $H\geq 50$.
Instead, in films grown with $p\leq0.7$, which have $\phi_E<0.6$ for all thicknesses,
the diffusivity increases with the thickness.

These different trends are balanced near $p=0.8$, in which $\phi_E\approx 0.6$ and $D_r\approx 0.3$.
The exact value $p^*$ in which this balance occurs depends on the choice of the thickness to be
compared with the thick film limit; for instance, if $H=50$ is chosen, the balance is estimated
to occur at $p^*=0.84$.

\subsection{Distributions of local solute currents}
\label{velocity}

\begin{figure}[!ht]
\center
\includegraphics[clip,width=0.23\textwidth,angle=0]{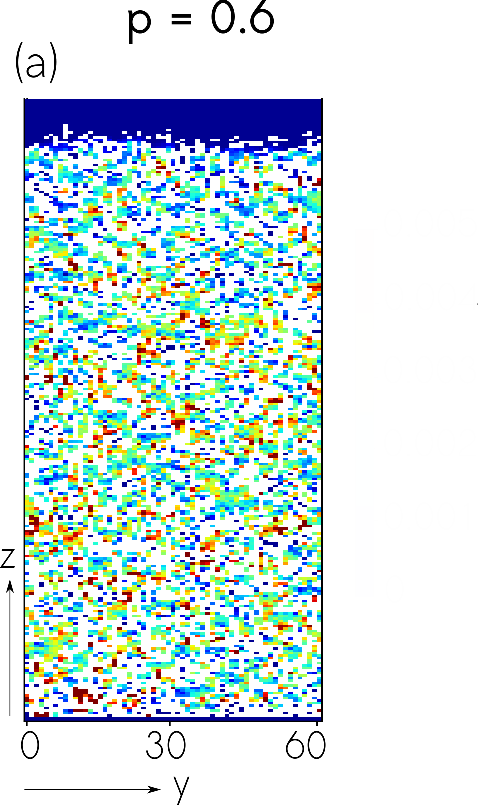} 
\includegraphics[clip,width=0.23\textwidth,angle=0]{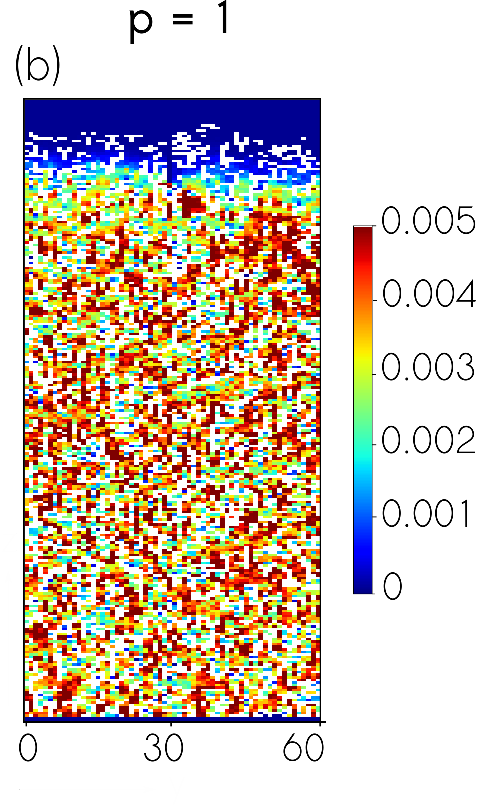}  

\vspace{0.1cm}

\includegraphics[clip,width=0.23\textwidth,angle=0]{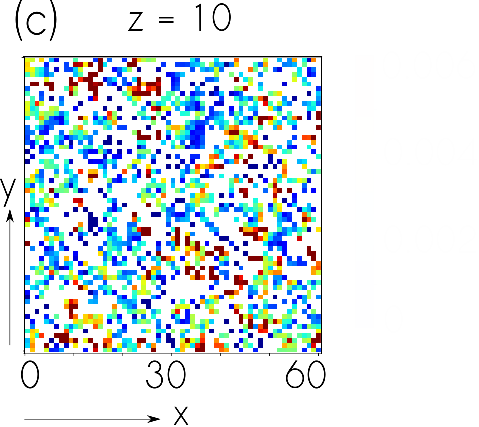}
\includegraphics[clip,width=0.23\textwidth,angle=0]{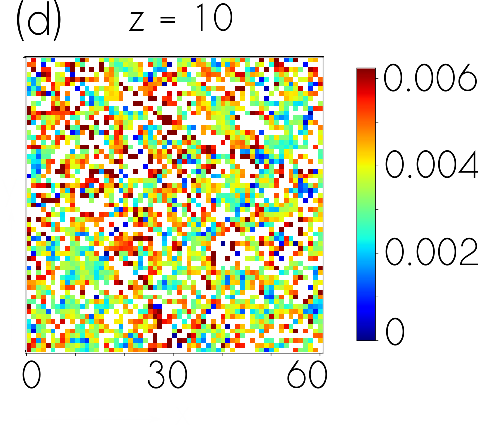}

\vspace{0.1cm}

\includegraphics[clip,width=0.23\textwidth,angle=0]{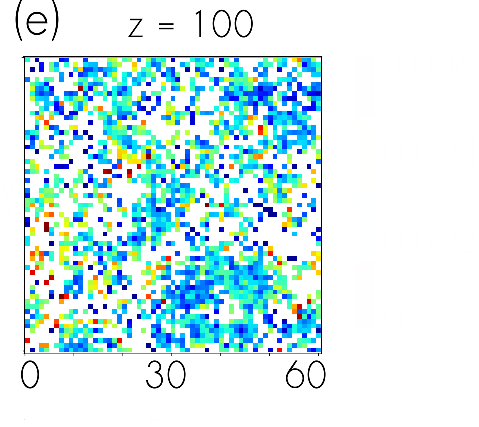}
\includegraphics[clip,width=0.23\textwidth,angle=0]{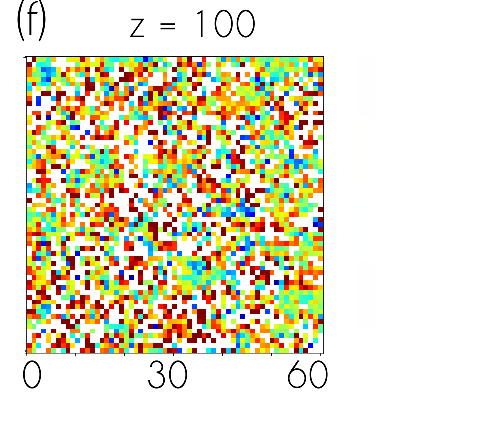}
\caption{
Configurations of the absolute value of the local current in (a),(b) vertical cross sections and
(c)-(f) horizontal cross sections.
The growth parameter $p$ and the heights of the horizontal sections are indicated.
In (a) and (b), the vertical scale is reduced by a factor $2$ for better visualization.
}
\label{fig_velocities}
\end{figure}

Figs. \ref{fig_velocities}(a) and \ref{fig_velocities}(b) show configurations of the absolute value
of the local solute current in vertical cross sections of films with thicknesses $\sim200$
grown with $p=0.6$ and $p=1$, respectively.
Figs. \ref{fig_velocities}(c)-(f) show configurations in horizontal cross sections
of the same films at different heights.
In films grown with $p=0.6$, the density of spots of large current ($j\gtrsim 0.005$; reddish color)
decreases as $z$ increases, while spots of intermediate current ($0.002\leq j\leq 0.003$)
become dominant at $z=100$.
The opposite trend is observed in the films grown with $p=1$,
in which the density of spots of large current increases with $z$.

\begin{figure}[!ht]
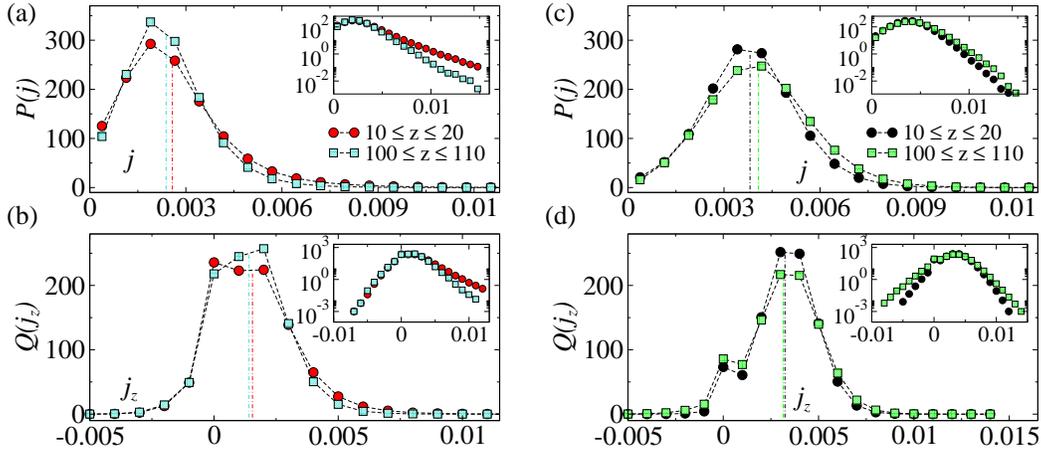

\center
\includegraphics[clip,width=0.4\textwidth,angle=0]{distrib_ab.eps} \quad
\includegraphics[clip,width=0.4\textwidth,angle=0]{distrib_cd.eps} \quad
\caption{
Probability density functions $P\left(j\right)$ of the absolute value of the local current
and $Q\left(j_z \right)$ of the $z$ component of the local current,
for films grown with (a),(b) $p=0.6$ and (c),(d) $p=1$ in two height intervals.
Dashed lines connecting the points are drawn to guide the eye.
The average values $\langle j\rangle$ and $\langle j_z\rangle$ are indicated by vertical dash-dotted
lines with the same colors of the corresponding data points.
The insets show the distribution tails in loglinear scale.
}
\label{distrib}
\end{figure}

Figs. \ref{distrib}(a) and \ref{distrib}(b) show the distributions $P\left(j\right)$
and $Q\left(j_z\right)$, respectively, in films grown with $p=0.6$ and two height intervals:
$10 \leq z \leq 20$ (lowest points) and $100 \leq z \leq 110$ (highest points).
The averages and the coefficients of variation of those distributions are presented in
Table \ref{tabelav}.
As $z$ increases, the peaks of the distributions increase, the average currents decrease,
and the relative widths decrease.
The decrease of $\langle j_z\rangle$ is directly related with the increase
in the effective porosity because the total current is the same in all horizontal sections.
The decrease in $c_{jz}$ means that a more homogeneous current distribution is attained at the
highest points.
The insets of Figs. \ref{distrib}(a)-(b) show magnified zooms of the distribution tails in log-linear scale;
they have faster decay at the highest points, in agreement with the visual observation of a smaller
number of large current spots in Figs. \ref{fig_velocities}(c) and \ref{fig_velocities}(e).

Figs. \ref{distrib}(c) and \ref{distrib}(d) show the same distributions for $p=1$ and
the corresponding averages and coefficients of variation are shown in Table \ref{tabelav}.
At the highest point, the distributions are wider and their peaks are smaller if
compared to the distributions at the lowest point;
the magnified zooms in the insets of Figs. \ref{distrib}(c)-(d) confirm that the tails have
slower decays at the highest points.
The inset of Fig. \ref{distrib}(d) also shows that the
fraction of sites with negative $j_z$ is larger at the highest points;
in those sites, the solute is forced to move in a direction opposite
to the average flux to contour the obstacles.
Since these films are predominantly grown with BD, this result is interpreted as an effect of the 
frequent lateral aggregation that blocks vertical paths for the flux.

\begin{table}[!ht]
\centering
\begin{tabular}{|c|c|c|c|c|}
\hline
                     & \multicolumn{2}{c|}{$p=0.6$}                & \multicolumn{2}{c|}{$p=1$}                  \\ \hline
                     & $10 \leq z \leq 20$ & $100 \leq z \leq 110$ & $10 \leq z \leq 20$ & $100 \leq z \leq 110$ \\ \hline
$\langle j\rangle$   & $0.00258$           & $0.00238$             & $0.00381$           & $0.00408$             \\ \hline
$c_j$                & $0.639$             & $0.538$               & $0.379$             & $0.399$               \\ \hline
$\langle j_z\rangle$ & $0.00155$           & $0.00141$             & $0.00324$           & $0.00315$             \\ \hline
$c_{jz}$             & $1.080$             & $1.022$               & $0.499$             & $0.603$               \\ \hline
\end{tabular}
\caption{
Average local currents $\langle j\rangle$ and $\langle j_z\rangle$ and 
coefficients of variation $c_j$ and $c_{jz}$ of the distributions in two height intervals, for films
grown with $p=0.6$ and $p=1$.
}
\label{tabelav}
\end{table}

\section{Discussion}
\label{discussion}

Here we discuss the relation between macroscopic structural and transport properties
(Sec. \ref{tortuosity}), the microscopic interpretation of the nontrivial behavior
of the effective diffusion coefficient (Sec. \ref{interpretation}),
the empirical relation between diffusivity and porosity (Sec. \ref{diffusionporosity}),
and possible relations with experimental realizations (Sec. \ref{relation}).

\subsection{Growth, structure, and diffusion}
\label{tortuosity}
    
In applications of porous media, it is expected that an increase in the porosity 
favors the diffusivity.
This is usually acomplished with changes in the growth conditions.
Our model confirms that it is possible with deposition techniques
if the relative rate of lateral particle aggregation is increased
(i.e. increase of $p$).
However, we also showed that larger porosity and larger diffusivity may be obtained
without changing the growth conditions, but increasing the film thickness,
for films typically with porosity below $0.6$.
The opposite trend is observed in the films grown with the largest rates of lateral aggregation,
which typically have porosity between $0.6$ and $0.7$ (considering small or large thicknesses).

The tortuosity $\tau$ is useful to quantify and interpret these types of structure-transport
relashionships.
Here we consider the empirical definition \citep{cussler}
\begin{equation}
D_r = \frac{\epsilon_E}{\tau} .
\label{deftau}
\end{equation}
In the thick film limit, it is generalized as $D_\infty =\epsilon_\infty /\tau$.
In Eq. (\ref{deftau}), $\epsilon_E$ is interpreted as the main factor to reduce the diffusive
current because it reduces the cross sectional area for solute flux;
for instance, if the medium has
straight channels connecting the two frontiers, we obtain $D_E={\epsilon_E}D_0$ and $\tau=1$.
However, if the channels have disordered shapes, the tortuosity represents the additional
effects of tortuous paths that the solute is forced to follow to cross the porous medium.

\begin{figure}[!ht]
\center
\includegraphics[clip,width=0.4\textwidth,angle=0]{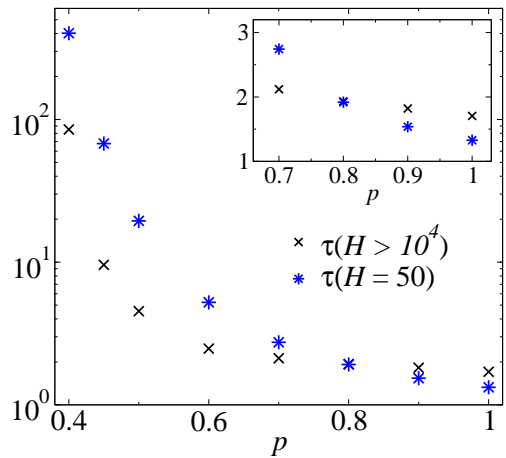}
\caption{
Tortuosities calculated in thin films with $H=50$ and in very thick films as a function of $p$. 
The inset highlight the data for $p\geq 0.7$.
}
\label{figtortuosity}
\end{figure}

Fig. \ref{figtortuosity} shows the tortuosities of thin films with $H=50$ and of the thick film limit
as function of $p$.
It confirms that the structure of the porous films have significant changes
when (i) the growth conditions vary (i.e. $p$ varies) and the thickness is constant
and (ii) the growth conditions are maintained and the thickness varies.
The inset of Fig. \ref{figtortuosity} highlights the region with large $p$.

If films with a constant thickness are compared, the tortuosity increases when $p$ decreases,
i.e. when the relative rate of film compaction increases.
Films with smaller thicknesses are more sentitive to changes in the growth conditions:
in thin films with $H=50$, the tortuosity varies from $1.3$ to $440$ as
the porosity decreases from $0.66$ to $0.1$;
in the thick film limit, a smaller variation is observed, from $1.7$ to $85$, with
a change in the porosity from $0.70$ to $0.19$.

When the growth conditions are maintained and the thickness varies, the tortuosity 
shows the transition between the two kinetic regimes at $p\approx 0.8$,
in which the effective porosity is in the range $0.58-0.62$.

In films grown with $p\geq 0.9$, whose porosity range is $0.62-0.70$,
$\tau$ is always between $1$ and $2$, and it increases with the thickness.
The maximal change in $\tau$ is near $25\%$, which indicates small changes in the pore structure.
This is confirmed by the small porosity ratio $r_\epsilon$ [Fig. \ref{figporosity}(b)].
These films also have small values of the roughness ratio $r_W$ (Fig. \ref{roughness}),
which shows that the outer surface (where porosity begins to form) evolves slowly as the film grows.
We understand that the inherent randomness of the deposition process increases the disorder
of the pore system as the film grows, which favors the increase in the tortuosity,
and this trend is not compensated by the slow increase in the porosity.

The decrease of $\tau$ with the thickness is observed in films grown with $p\leq 0.7$,
whose porosities are below $0.58$.
From $H=50$ to the thick film limit, our simulations showed variations in
$\tau$ by factors ranging from
$\approx 1.3$ ($\epsilon_\infty= 0.57$) to $\approx 4.7$ ($\epsilon_\infty= 0.19$).
Thus, there are significant changes in the pore systems as these films grow.
Indeed, Figs. \ref{roughness} and \ref{figporosity}(b) show larger values of the
roughness ratio $r_W$ and of the porosity ratio $r_\epsilon$ in these conditions.
Our interpretation is that the faster widening of the pore system compensates the increasing disorder
imposed by the random nature of the deposition process.

\subsection{Microscopic interpretation}
\label{interpretation}
    
In films grown with $p=0.6$, 
Figs. \ref{distrib}(a)-(b) show that the local current distributions are narrower at the highest points.
This differs from the high porosity films grown with $p=1$, in which those distributions are wider
at the highest points [Figs. \ref{distrib}(c)-(d)].

When particles randomly move in a porous medium, the increase in the tortuosity
is related to the presence of obstacles to their random motion.
The delays caused by the disorder may be expressed in terms of distributions of
waiting times (or hopping times).
An increase in the structural disorder is usually associated with an increase in the dispersion
of waiting times.
For instance, in the case of fractal networks, the self-similar pore size distributions lead to
subdiffusion \citep{havlin}, i.e. time-decreasing diffusion coefficients;
the corresponding distributions of waiting times are also self-similar, which inspires the
approaches of continuous time random walks \cite{bouchaud,metzler2014}.
In the present model,
the dispersion of waiting times is translated into a dispersion of local currents.

This reasoning leads to a microscopic interpretation of the two kinetic regimes observed here.
The distributions in Figs. \ref{distrib}(a)-(b) show that the dispersion of waiting times 
is smaller at higher points.
Thus, the evolution of the local structure with the height is favorable for the increase of the
diffusivity.
These results, which were shown for films grown with $p=0.6$, are representative of the other
films grown with $p\leq0.7$.
However, in the films with $p=1$, Figs. \ref{distrib}(c)-(d) show that the dispersion in the
waiting times increases with the height, so the local structure evolution is unfavorable
for the diffusion.
These results are representative of the films with high porosity, i.e. grown with $p\geq0.9$.
In summary, the decrease in the fluctuations of the local currents combined with the
increase in the local porosity are the microscopic features that correlate with the
thickness increasing diffusivity.

The macroscopic and the microscopic interpretations presented here are consistent
and are based on the calculated global and local quantities, respectively.
However, the location of the transition probability $p^*$ between the two kinetic regimes
cannot be predicted exactly (or with high accuracy) from the available data.

\subsection{Relation between diffusion coefficient and porosity}
\label{diffusionporosity}

Works on diffusion in porous media frequently obtain empirical power law relations
as $D_r={\epsilon_E}^m$, with positive exponents $m$ \citep{dullien,Adler,grathwohl}.
This type of relation is known as the Archie law \citep{archie1942}.
Here we check whether this law is applicable to the films grown with the BD-RDSR model,
considering the quantities obtained in the thick film limit.

\begin{figure}[!th]
\center
\includegraphics[clip,width=0.4\textwidth,angle=0]{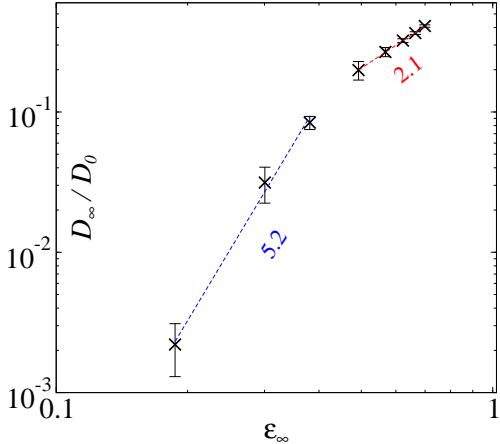}
\caption{ 
Diffusion coefficient as a function of the porosity in the thick film limit.
Linear fits of the data with $0.5\lesssim\epsilon_\infty\lesssim0.7$ and with $\epsilon_\infty\lesssim0.4$,
and the corresponding slopes are shown.
}
\label{archie}
\end{figure}

Fig. \ref{archie} shows a bilogarithmic plot of $D_\infty$ versus $\epsilon_\infty$.
It has a downward curvature and, consequently, does not allow a reliable fit of the data
in the whole range of porosity studied here.
In other words, the empirical Archie law is not applicable to the complete set of porous media
generated by the BD-RDSR model.

We also tried to fit different parts of the plot in Fig. \ref{archie}.
For the largest porosities, $0.5\lesssim\epsilon_\infty\lesssim0.7$, 
the fit has a slope $2.05\pm 0.02$, which is close to the values obtained in
many natural materials \citep{grathwohl}.
However, this fit cannot be extrapolated because it deviates $\approx 10\%$ from the free medium value,
$D_\infty=1$ for $\epsilon_\infty=1$; thus, even in this restricted range of porosity,
the Archie relation cannot be consistently used.
For the smallest porosities, $\epsilon_\infty <0.4$, the porous media are closer to the critical
percolation point and the fit has slope $5.2\pm 0.3$.
A scaling approach for the diffusion in off-critical conditions
\citep{havlin,havlin1983b} predicts a slope $4.8\pm 0.2$, which is close to our estimate;
see details in Sec. SI.VI of the Supporting Information.
However, this should not be interpreted as an application of the Archie law because the fit is
also inconsistent with the free medium diffusivity.

\subsection{Possible relations with experimental works}
\label{relation}

The following discussion has a focus on battery electrodes because
it was already shown that ion diffusion in their pores frequently is the process that
limits the battery performance \citep{gao2018,hossain2019}.
Moreover, experiments and modeling show that adjusting the porosity and the tortuosity
by varying the thickness may be a key step to improve electrode efficiency
\citep{dai2016,colclasure2020}.

In a recent work \citep{suACSAMI2020}, a 13.5 $\mu$m thick pristine carbon porous matrix
used in Li-O${}_2$ batteries
was shown to have porosity $0.19$ and tortuosity $29.97$, while a discharged cathode
had the respective values $0.33$ and $4.80$.
Considering the carbon atomic size, the comparison with our model must consider the thick film limit,
in which we obtained $\tau=4.50$ for $\epsilon_{\infty}=0.38$
and $\tau=85$ for $\epsilon_{\infty}=0.19$;
the first pair of model values is close to the experimental one for the discharged cathode.
Another recent work \citep{vierrath2015} presented a carbon binder domain (CBD) of a LiCoO$_2$ battery
with thickness of a few micrometers, porosity $0.58$, and tortuosity $2.3-3.5$.
For this porosity, our model gives $\tau\approx2.1$, which is also close to the experimental estimates.
Finally, a study of calendered, negative electrodes ($\sim 100\mu$m thick)
without CBD show porosities $0.35-0.40$ and tortuosities $4.9-5.9$
\citep{viretta2018}.
Interpolation of the model data in Fig. \ref{figtortuosity} gives $\tau=4.2--6.5$
in the same porosity range. 
Thus, despite the strong approximations of the BD-RDSR model, it leads to relations between porosity
and tortuosity that are quantitatively close to experimental values.

We are not aware of processes in which the growth conditions were maintained and
an increase in the diffusivity was observed.
However, in growth of composition graded films, the physical conditions were dynamically changed to
increase the porosity and to decrease the tortuosity with the thickness \citep{liuJES2017,cheng2020}.
Despite that difference, our results may be helpful to applications using this technique;
for instance, if the porosity formation is controlled by the adsorption,
a rapid increase of the roughness with the thickness may anticipate
a favorable evolution of the pore system.

On the other hand, in a recent study with reduced graphene oxide (ErGO) electrodes produced 
by electrodeposition,
the electrolyte diffusivity was shown to decrease nearly $4$ times as the film thickness increases
from $3\mu$m to $79\mu$m \citep{zhan2020}.
In this case, it is possible that the deposition kinetics is (at least partially) responsible for
the decrease in the diffusivity, as in the regime of thickness-hampered transport of our model.
If the electrodeposition conditions can be controlled in this system, we believe that it
is a potential candidate for an experimental search of the transition to a
regime of thickness-facilitated transport, as suggested by the model.

In any application, the benefits or disadvantages of the results presented here have to
be correlated with other effects of increasing a film thickness.
For energy storage applications, larger thicknesses are usually favorable
to increase the areal energy density.
Thus, if the regime of thickness-facilitated transport (medium to low porosities in our model)
is applicable, the beneficial effect to ion diffusion is an additional reason to deposit thicker films.
However, in the regime of thickness-hampered transport, the increase in the tortuosity with the
thickness is relatively small ($\lesssim 25\%$), so this should not be a major concern in comparison
with the energetic advantage of thicker films.
Thus, whenever the deposition conditions are consistent with the approximations of the present
model, we generally expect that the increase of thickness is helpful for energy storage applications,
even if there is a small loss in diffusivity.

At this point, it is important to highlight
the physical conditions considered in the model studied here:
the particle flux towards the deposit must be (at least partially) collimated, the relaxation
of deposited particles is of short range, and subsurface relaxation during the growth is negligible.
Alternatively, the relaxation mechanism could consider preferential aggregation at the sites
with the largest numbers of NN bonds instead of the sites with the lowest heights, for instance
using the models of Das Sarma and Tamborenea \citep{dt} and of Wolf and Villain \citep{wv}.
However, there is numerical evidence that these two models are in the same growth class (EW)
of RDSR in 3D \citep{chamereis2004}.
Thus, mixing the BD rules and the rules of one of those models will also lead to a competitive
KPZ-type model, in which the porosity is expected to increase with the thickness for a fixed $p$.
The thickness dependence of the diffusion coefficient cannot be predicted from these properties,
but the above similarities suggest that the dependence may be the same as that of BD-RDSR.
On the other hand, if the mobility of the adsorbed particles is large and they can eventually fill
distant pores, then more drastic changes in the film structure are expected and the conclusions
drawn from BD-RDSR may fail.

\section{Conclusion}
\label{conclusion}

We studied the relation between the growth kinetics of porous films, their surface and bulk
structures, and the diffusive transport of a solute in their pores.
We used a deposition model in a lattice in which a single parameter
controls the relative rates of lateral aggregation of incident particles (responsible for pore formation)
and of relaxation after deposition (responsible for compaction of the deposits).
When the deposition conditions change by increasing this rate, we always observe
the increase of the growth velocity, of the surface roughness,
of the effective porosity, and of the diffusivity.

We also compared films produced in the same growth conditions but with different thickness.
As the thickness increases, the effective porosity slowly increases; this is the same trend predicted for
the total porosity by the KPZ theory of interface roughening.
In films with porosity below $0.58$, the effective diffusion coefficient increases with the thickness.
This is partly due to the increase of porosity, but mostly due the decrease of the tortuosity.
For instance, comparing films whose thicknesses correspond to $50$ and to $\geq{10}^4$
layers of particles,
the tortuosity may be reduced by a factor $\sim 5$ while the porosity increases by a factor $\sim 2$.
At microscopic level, we observed that the distributions of local solute currents in these
deposits are narrower at higher points, i.e. the currents are more uniform.
This indicates smaller dispersion in the waiting times
of the diffusing species and qualitatively explains the thickness-facilitated transport.
In films with porosities $\sim 0.65-0.7$
(the maximal values obtained with the model),
the tortuosity decreases with the thickness, but the maximal changes are near $25\%$.
This regime of thickness-hampered transport is characterized by larger widths of
the local current distributions at higher points.

The values of porosity and tortuosity in thick films are close
to the values obtained in porous electrodes produced with different techniques
\citep{suACSAMI2020,vierrath2015,viretta2018}.
Thus, despite the simple stochastic rules of our model,
it may be reasonable to predict qualitative relations between
growth, structure and diffusive transport in porous materials of technological interest.
Quantitative relations are expected to be predicted by simulations of improved models
which account for details of particular growth processes.
Possible extensions may consider the effects of non-collimated particle flux and longer
diffusion lenghts of the adsorbed species, as shown in recent models that reproduce
morphological properties of dendritic films \citep{aryanfarJCP2015,reiscapriotaleb2017,vishnugopi2020}.
In these cases, the dendrite orientations and the porosity variations that are expected
due to the growth instabilities should affect the pore diffusivity.

We also believe that the results presented here or extensions of our approach
may be useful to study porous media other than electrodes,
and of interest in different areas.
For instance, diffusion simulations were recently performed
in models of deposited porous media proposed for sedimentary rocks \citep{giri2013},
but no thickness effect was analyzed.
From a point of view of fundamental science,
the transition predicted between kinetic regimes of thickness-facilitated and thickness-hampered
transport may be of interest both for experimental investigation
and for mathematical treatment.

\begin{suppinfo}

The Supporting Information file shows details of the simulations, the method to calculate the
effective diffusion coefficients, the evolution of the growth velocity, the extrapolation of
the porosity to the thick film limit, the variations of the numbers of connected domains,
and the scaling approach for diffusion coefficient in low porosities.

\end{suppinfo}

\section*{Acknowledgment}
FDAAR acknowledges support from the Brazilian agencies CNPq (305391/2018-6), FAPERJ 
(E-26/110.129/2013, E-26/202.881/2018), and CAPES (88887.310427/2018-00 - PrInt).
GBC acknowledges support from CAPES (88887.198125/2018-00).
RA acknowledges support from CAPES (88887.370801/2019-00 - PrInt). 

\newpage

\begin{longtable}{|p{0.1\textwidth}|p{0.7\textwidth}|p{0.15\textwidth}|}
\caption{List of symbols used in this work.}\\ \hline
Symbol & Quantity & Dimension \\ \hline
 $a$ & length of a site edge & L \\ \hline
 $c_j$ & coefficient of variation of $P\left(\vec{j}\right)$ & dimensionless \\ \hline
 $c_{jz}$ & coefficient of variation of $Q\left(j_z\right)$ & dimensionless \\ \hline
 $C$ & concentration of the solute in a pore site & dimensionless \\ \hline
 $\overline{C}$ & average steady state solute concentration & dimensionless \\ \hline
 $D_{E}$ & effective diffusion coefficient of the solute in the porous medium & L${}^{2}$T${}^{-1}$ \\ \hline
 $D_{0}$ & diffusion coefficient of the solute in a free medium & L${}^{2}$T${}^{-1}$ \\ \hline
 $D_{r}$ & normalized or relative effective diffusion coefficient & dimensionless \\ \hline
 $D_\infty$ & effective diffusion coefficient in the thick film limit & dimensionless \\ \hline
 $h(x,y)$ & maximum height of a particle in a column $(x,y)$ of the deposit & L \\ \hline
 $H$ & thickness of the deposits & dimensionless \\ \hline
 $\vec{j}$  & local average solute current & dimensionless \\ \hline
 $\vec{j_s}$ & local average solute current & L${}^{-2}$T${}^{-1}$ \\ \hline
 $j_z$ & $z$ component of the dimensionless local average solute current & dimensionless \\ \hline
 $\langle j\rangle$ & average of the absolute value of $\vec{j}$ & dimensionless \\ \hline
 $\langle j_z\rangle$ & average of $j_z$ & dimensionless \\ \hline
 $J_{in}$ & average solute current that crosses the substrate & L${}^{-2}$T${}^{-1}$ \\ \hline
 $J_{out}$ & average solute current that crosses the top outer surface & L${}^{-2}$T${}^{-1}$ \\ \hline
 $J_{s}$ & stationary current (averaging $J_{in}$ and $J_{out}$) & L${}^{-2}$T${}^{-1}$ \\ \hline
 $L$ & lateral size of the deposition substrate & L \\ \hline
 $m$ & cementation exponent of Archie law & dimensionless \\ \hline
 $N_d$ & total number of deposited particles & dimensionless \\ \hline
 $p$ & probability that a particle aggregates according to the ballistic deposition rule & dimensionless \\ \hline
 $p_c$ & critical percolation probability & dimensionless \\ \hline
 $p^{*}$ & probability of transition between regimes of thickness-dependent diffusivity & dimensionless \\ \hline
 $P$ & probability density function of the absolute value of $\vec{j}$ & dimensionless \\ \hline
 $Q$ & probability density function of $j_z$ & dimensionless \\ \hline
 $r_W$ & ratio between the roughness of thick films and of thin films & dimensionless \\ \hline
 $r_\epsilon$ & ratio between the porosity of thick films and of thin films & dimensionless \\ \hline
 $t_g$ & growth time of the deposits & dimensionless \\ \hline
 $v_a$ & growth velocity of the deposits & dimensionless \\ \hline
 $v_g$  & instantaneous growth velocity of the deposits & dimensionless \\ \hline
 $v_\infty$ & long time instantaneous growth velocity of the deposits & dimensionless \\ \hline
 $V_c$ & total volume of the connected pore system & L${}^3$ \\ \hline
 $V_T$ & total volume of the deposits & L${}^3$ \\ \hline
 $W$ & roughness of the deposits & dimensionless \\ \hline
 $\beta$ & growth exponent of film roughness & dimensionless \\ \hline
 $\Delta_\epsilon$ & relative difference between the total and the effective porosities & dimensionless \\ \hline
 $\epsilon_T$ & total porosity of the deposits & dimensionless \\ \hline
 $\epsilon_E$ & effective or connected porosity of the deposits & dimensionless \\ \hline
 $\epsilon_\infty$ & effective porosity extrapolated in the thick film limit & dimensionless \\ \hline
 $\nu$ & rate at which solute molecules execute random walks & T${}^{-1}$ \\ \hline
 $\tau$ & tortuosity of the porous medium & dimensionless \\ \hline
 $\xi_I$ & intrinsic roughness & dimensionless \\ \hline
 \label{NomenclatureTable}
 \end{longtable}


\providecommand{\latin}[1]{#1}
\providecommand*\mcitethebibliography{\thebibliography}
\csname @ifundefined\endcsname{endmcitethebibliography}
  {\let\endmcitethebibliography\endthebibliography}{}

\newpage




\pagebreak
\thispagestyle{empty}

\begin{center}
\vspace{20cm}

    \title{\huge{Supporting Information}\\ \vskip 1cm \Large\textbf{{Effects of the Growth Kinetics on Solute Diffusion in Porous Films}}}\\[.5cm]
    Gabriela B. Correa$^{*}$, 
    Renan A. L. Almeida, 
    and F\'abio D. A. Aar\~ao Reis$^{\ddag}$\\[.1cm]
    
    {\itshape Instituto de F\'{i}sica, Universidade Federal Fluminense, Avenida Litor\^{a}nea s/n, 24210-340 Niter\'{o}i, RJ, Brazil\\}
    E-mail: $^{*}$gabriela\_barreto@id.uff.br; 
    $^{\ddag}$reis@if.uff.br\\

\vspace{5cm}
\end{center}
\newpage

\setcounter{equation}{0}
\setcounter{section}{0}
\setcounter{figure}{0}
\setcounter{table}{0}
\setcounter{page}{1}
\renewcommand{\theequation}{S\arabic{equation}}
\renewcommand{\thefigure}{S\arabic{figure}}
\renewcommand{\bibnumfmt}[1]{[S#1]}
\renewcommand{\citenumfont}[1]{S#1}

\renewcommand\thesection{SI.\Roman{section}}
\renewcommand\thefigure{S\arabic{figure}}
\renewcommand\theequation{S\arabic{equation}}
\renewcommand\thepage{S\arabic{page}}

\section{Simulations of film growth and calculation of structural properties}
\label{simulationdetails}

The calculation of average thickness, roughness, and total porosity depends only on 
the configuration of the top outer surface. 
For several values of $p$, those quantities were averaged in $300$ film configurations 
with lateral size $L=512a$ and grown up to $t_g\approx {10}^3$, which provides results with
accuracy $\lesssim 1\%$.

The enumeration of connected porous domains and the calculation of the effective porosity
requires knowledge from the whole film configuration between the two frontiers.
Counting of connected domains was performed by 
a three-dimensional Hoshen-Kopelman code \cite{hoshen}.
Average quantities were obtained out of 300 samples for lateral size $L = 512a$
and out of 1000 samples for lateral sizes $L = 128a$ and $256a$.
Apart from $p = 0.4$, neither appreciable finite-size nor finite-sampling effects were observed.


\section{Simulations of diffusive transport}
\label{diffusionsimulation}

\begin{figure}[!ht]
\center
\includegraphics[clip,width=0.4\textwidth,angle=0]{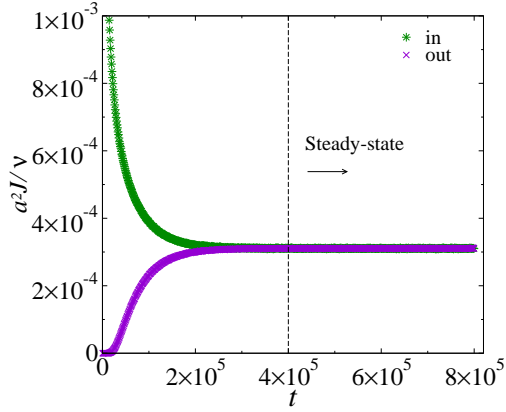}
\caption{
Time evolution of the scaled currents flowing into and out of the films grown
with $p=1$ up to $t_g=70$ (thickness $H \approx 225$ and total porosity $\epsilon_T\approx 0.69$).
}
\label{currents}
\end{figure}

Simulation of infiltration begins with no solute in the pores.
We measure the solute currents $J_{in}$ at the substrate ($z=0$) and $J_{out}$ 
near the outer surface ($z/a = H$).
These currents are  defined as the numbers of molecules flowing into and out of the
pore system, respectively, per unit flat area $L^2$ and unit time.
After some time, the system reaches a steady state, in which these two currents fluctuate
around the same value.
This is illustrated in Fig. \ref{currents}.

In a medium with thickness $H$, the average distance between the source and the drain is $H+1$.
In steady state transport, the average value of the current, $J_s$,
is related to the effective diffusion coefficient of the solute, $D_E$, 
by $J_s=D_E/\left[ \left( H+1\right) a^4\right]$ (first Fick law).
In a three dimensional free medium, the diffusion coefficient is $D_0=\nu a^2/6$.
The effective diffusion coefficient is consequently given by Eq. (6) of the main text,
The dimensionless current $J_s a^2/\nu$ is directly determined in the
simulations, as illustrated in Fig. \ref{currents}.
In any disordered system, we expect that $D_E/D_0<1$.

An alternative method to probe the effect of a disordered medium on diffusion processes
is the release of non-interacting random walkers, or tracers, in that medium.
To avoid effects of the finite system size (including the thickness), the tracers are released
far from the borders and the choice of the maximal walking time has to warrant a negligible
probability of reaching those borders.
However, our aim is to model systems with mass flux across the sample and to determine thickness effects.
In this case, the approach described above is more appropriate than tracer diffusion
simulations.

The simulation of transport is the most time consuming part of our numerical work.
For each configuration of the film (grown with a given $p$), the calculation has to be separately
performed in several thicknesses and, for each thickness, convergence to the steady state is
necessary.
Reaching the steady state takes a time of order $H^2/\nu$ in the samples with higher porosity
($\sim 70\%$), but may be two orders of magnitude slower in tortuous media
with low porosity ($\lesssim 20\%$).
In the timescale $1/\nu$, it is necessary to update the positions of 
a number of solute molecules of order $\epsilon_E{\left( L/a\right)}^2 H$,
so the total number of updates
in a given simulation is, at least, of order $\epsilon_E {\left( L/a\right)}^2 H^3$.

For these reasons, most simulations of diffusive transport were performed in lateral sizes
$L/a=64$ and $128$ and thicknesses $H\approx 300$ or smaller.
For each $p\geq 0.45$ and $20$-$30$ values of the thickness $H$,
$40$ film configurations were generated for these calculations.
The comparison of results in sizes $L/a=64$ and $128$ showed weak effects of the finite size $L$.
Some films with smaller thicknesses were also simulated in size $L=256$ to confirm this trend.
For $p=0.4$, finite-size effects in $L/a=128$ were observed because these films are closer to
the critical percolation point; consequently, we performed simulations in $L/a=256$.


\section{Evolution of the thickness}
\label{thicknessevolution}

\begin{figure}[!ht]
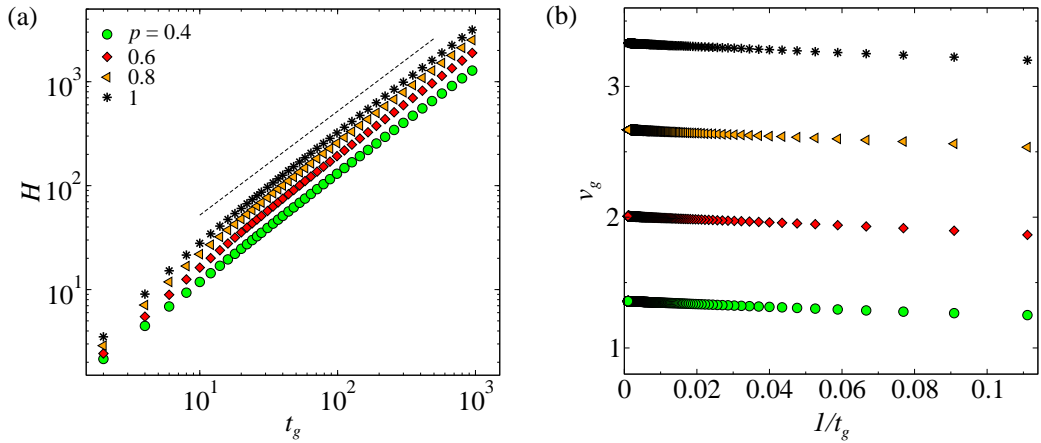

\center
\includegraphics[clip,width=0.4\textwidth,angle=0]{thickness_a.eps} \quad
\includegraphics[clip,width=0.4\textwidth,angle=0]{thickness_b.eps}

\caption{
(a) Average film thickness as function of the growth time for several values of $p$.
The uncertainties are smaller than the size of the symbols.
(b) Growth velocity estimated in time intervals $\Delta t_g=2$ as a function of the reciprocal of the
growth time for several values of $p$.
}
\label{thickness}
\end{figure}

Fig. \ref{thickness}(a) shows the time evolution of the thickness for $p\geq 0.4$.
The good linearity observed at long times indicate that the growth velocities converge
to constant values in large thicknesses.
Fig. \ref{thickness}(b) shows the dimensionless growth velocity $v_g$
[Eq. (7) of the main text] as a function of $1/t_g$, which facilitates the observation of
the time increase of $v_g$.

\section{Extrapolation of the porosity to the thick film limit}
\label{extrapolationporosity}

\begin{figure}[!ht]
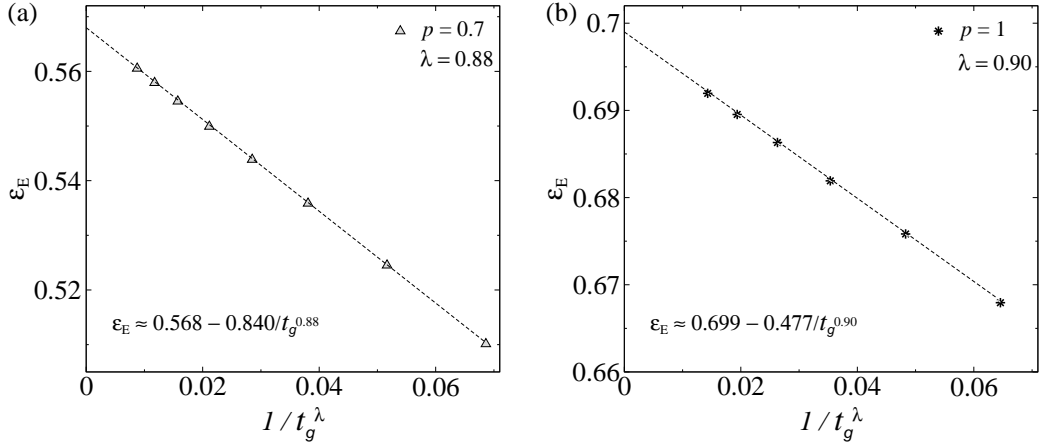

\centering
\includegraphics[clip,width=0.4\textwidth,angle=0]{extrap_a.eps} \quad
\includegraphics[clip,width=0.4\textwidth,angle=0]{extrap_b.eps}
\caption{ Effective porosity as function of $1/t_g^{\lambda}$ for (a) $p=0.7$ and (b) $p=1$, with the values of $\lambda$ indicated in each plot. Dashed lines are linear fits of the data.}
\label{extrapP}
\end{figure}

Based on Eq.(9) of the paper, 
we propose similar scaling relation for the effective porosity:
\begin{equation}
\epsilon_E \approx \epsilon_\infty - \frac{B}{t_g^{\lambda}} ,
\label{porEscaling}
\end{equation}
where $\epsilon_\infty$, $B$, and $\lambda$ are positive constants.
Figs. \ref{extrapP}(a) and \ref{extrapP}(b) show $\epsilon_E$ as a function of $1/t_g^{\lambda}$
for $p=0.7$ and $p=1$, respectively, using the values of $\lambda$ that provide the best
linear fits in each case.

The fits to Eq. (\ref{porEscaling}) provide estimates of $\epsilon_\infty$,
which are the effective porosities in the thick film limit.
In all cases, this is expected to be the effective porosity of films with
$H\geq {10}^4$ with accuracy $\lesssim 1\%$.


\section{Number of connected domains}
\label{numberdomains}

\begin{figure}[!ht]
\centering
\includegraphics[clip, width = 0.4\textwidth,angle=0]{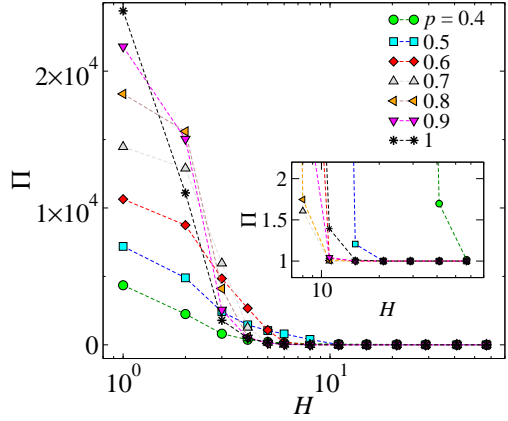}
\caption{Average number of connected porous domains as function of the average thickness.
Results are obtained in size $L = 512a$ and averaging over 300 samples.
The inset shows a zoom of the region near $\Pi = 1$. }
\label{connecteddomains}
\end{figure}

Fig. \ref{connecteddomains} shows the average number of connected porous domains $\Pi$ for several values of $p$.
For $0.5 \leq p \leq 1$, a single domain is observed in films with $H\gtrsim 20$;
for $p=0.4$, this is observed with $H\gtrsim 60$.

\section{Diffusion coefficient in low porosities}
\label{lowporosity}

A check of the reliability of our calculations of diffusion coefficients is
their behavior for low effective porosity, in which the pore systems are close to the critical percolation 
point located at $p_c\approx 0.35$ \cite{yuamar2002}.
Let $\epsilon_c$ be the value of the total porosity at this critical point.
In the connected porous domain, a scaling approach \cite{havlin1983b,havlinAdvPhys2002} predicts that
the diffusion coefficient of a random walker scales as
$D_T\sim {\left( \epsilon_T-\epsilon_c\right)}^{\tilde{\mu}\nu_p-\beta_p}$,
where $\tilde{\mu}$ is termed conductivity exponent and $\nu_p$ and $\beta_p$ stand for percolation
exponents.
Here, the total pore volume fraction $\epsilon_T$ parallels the occupation
probability of lattice models with random distributions of occupied and empty sites
\cite{stauffer,havlinAdvPhys2002}.

The diffusion coefficient $D_T$ is reduced in comparison with the free medium value only due to the
tortuous paths of the porous medium, while the effective diffusion coefficient also decreases due
to the restricted pore volume in which the solute moves;
thus, they are related as $D_E=\phi_E D_T$.
In the notation of the thick film limit used here, if $D_T$ is normalized by the free medium
diffusivity, we have $D_\infty=\epsilon_\infty D_T$.

Basic percolation theory also sets that the effective porosity close to the
critical point scales as $\epsilon_\infty\sim {\left( \epsilon_T-\epsilon_c\right)}^{\beta_p}$ \cite{stauffer};
this is the relation that originally defines the exponent $\beta_p$.
Considering the scaling relations between percolation exponents \cite{stauffer,havlinAdvPhys2002},
we obtain
\begin{equation}
D_E\sim {\epsilon_\infty}^q \qquad , \qquad q=\frac{\tilde{\mu}}{ 3-d_f} ,
\label{defp}
\end{equation}
where $d_f$ is the fractal dimension of the critical percolation cluster.
The available numerical estimates $\tilde{\mu}=2.26\left( 4\right)$ \cite{normand1995} and
$d_f=2.53\left( 4\right)$ \cite{jan1998} give $q=4.8\left( 2\right)$,
which is close to the numerical estimate $5.2\pm 0.3$ shown in the main text.
Here, $q$ plays a role similar to the cementation exponent $m$ of the Archie law \cite{grathwohl},
but Eq. (\ref{defp}) does not fit the values of $D_E$ in large porosities.



\bibliographystyle{unsrt}


\begin{mcitethebibliography}{70}
\providecommand*\natexlab[1]{#1}
\providecommand*\mciteSetBstSublistMode[1]{}
\providecommand*\mciteSetBstMaxWidthForm[2]{}
\providecommand*\mciteBstWouldAddEndPuncttrue
  {\def\EndOfBibitem{\unskip.}}
\providecommand*\mciteBstWouldAddEndPunctfalse
  {\let\EndOfBibitem\relax}
\providecommand*\mciteSetBstMidEndSepPunct[3]{}
\providecommand*\mciteSetBstSublistLabelBeginEnd[3]{}
\providecommand*\EndOfBibitem{}
\mciteSetBstSublistMode{f}
\mciteSetBstMaxWidthForm{subitem}{(\alph{mcitesubitemcount})}
\mciteSetBstSublistLabelBeginEnd
  {\mcitemaxwidthsubitemform\space}
  {\relax}
  {\relax}

\bibitem[Li \latin{et~al.}(2017)Li, Du, Ruther, An, David, Hays, Wood, Phillip,
  Sheng, Mao, Kalnaus, Daniel, and {Wood III}]{liJOM2017}
Li,~J.; Du,~Z.; Ruther,~R.~E.; An,~S.~J.; David,~L.~A.; Hays,~K.; Wood,~M.;
  Phillip,~N.~D.; Sheng,~Y.; Mao,~C. \latin{et~al.}  Toward low-cost,
  high-energy density, and high-power density lithium-ion batteries. \emph{J.
  Occup. Med.} \textbf{2017}, \emph{69}, 1484--1496\relax
\mciteBstWouldAddEndPuncttrue
\mciteSetBstMidEndSepPunct{\mcitedefaultmidpunct}
{\mcitedefaultendpunct}{\mcitedefaultseppunct}\relax
\EndOfBibitem
\bibitem[Wang \latin{et~al.}(2020)Wang, Li, Hao, and Tan]{wangACSAEM2020}
Wang,~F.; Li,~X.; Hao,~X.; Tan,~J. Review and recent advances in mass transfer
  in positive electrodes of Aprotic {Li-O${}_2$} Batteries. \emph{ACS Appl.
  Energy Mater.} \textbf{2020}, \emph{3}, 2258--2270\relax
\mciteBstWouldAddEndPuncttrue
\mciteSetBstMidEndSepPunct{\mcitedefaultmidpunct}
{\mcitedefaultendpunct}{\mcitedefaultseppunct}\relax
\EndOfBibitem
\bibitem[Koros and Zhang(2017)Koros, and Zhang]{koros2017}
Koros,~W.~J.; Zhang,~C. Materials for next-generation molecularly selective
  synthetic membranes. \emph{Nat. Mater.} \textbf{2017}, \emph{16},
  289--297\relax
\mciteBstWouldAddEndPuncttrue
\mciteSetBstMidEndSepPunct{\mcitedefaultmidpunct}
{\mcitedefaultendpunct}{\mcitedefaultseppunct}\relax
\EndOfBibitem
\bibitem[Lin \latin{et~al.}(2019)Lin, Xiang, Li, Cui, Qian, Zhou, and
  Chen]{linCCRev2019}
Lin,~R.-B.; Xiang,~S.; Li,~B.; Cui,~Y.; Qian,~G.; Zhou,~W.; Chen,~B. Our
  journey of developing multifunctional metal-organic frameworks. \emph{Coord.
  Chem. Rev.} \textbf{2019}, \emph{384}, 21--36\relax
\mciteBstWouldAddEndPuncttrue
\mciteSetBstMidEndSepPunct{\mcitedefaultmidpunct}
{\mcitedefaultendpunct}{\mcitedefaultseppunct}\relax
\EndOfBibitem
\bibitem[{Santamar\'ia-Holek} \latin{et~al.}(2019){Santamar\'ia-Holek},
  {Hern\'andez}, {Garc\'ia-Alc\'antara}, and {Ledesma-Dur\'an}]{holek2019}
{Santamar\'ia-Holek},~I.; {Hern\'andez},~S.~I.; {Garc\'ia-Alc\'antara},~C.;
  {Ledesma-Dur\'an},~A. Review on the macro-transport processes theory for
  irregular pores able to perform catalytic reactions. \emph{Catalysis}
  \textbf{2019}, \emph{9}, 281\relax
\mciteBstWouldAddEndPuncttrue
\mciteSetBstMidEndSepPunct{\mcitedefaultmidpunct}
{\mcitedefaultendpunct}{\mcitedefaultseppunct}\relax
\EndOfBibitem
\bibitem[Zheng \latin{et~al.}(2018)Zheng, Geng, Zhang, Chen, and
  Li]{zhengChemA2018}
Zheng,~Y.; Geng,~H.; Zhang,~Y.; Chen,~L.; Li,~C.~C. Precursor-based synthesis
  of porous colloidal particles towards highly efficient catalysts. \emph{Chem.
  Eur. J.} \textbf{2018}, \emph{24}, 10280--10290\relax
\mciteBstWouldAddEndPuncttrue
\mciteSetBstMidEndSepPunct{\mcitedefaultmidpunct}
{\mcitedefaultendpunct}{\mcitedefaultseppunct}\relax
\EndOfBibitem
\bibitem[Heubner \latin{et~al.}(2020)Heubner, Langklotz, {L\"ammel}, Schneider,
  and Michaelis]{heubner2020}
Heubner,~C.; Langklotz,~U.; {L\"ammel},~C.; Schneider,~M.; Michaelis,~A.
  Electrochemical single-particle measurements of electrode materials for
  {Li-ion batteries: Possibilities,} insights and implications for future
  development. \emph{Electrochim. Acta} \textbf{2020}, \emph{330}, 135160\relax
\mciteBstWouldAddEndPuncttrue
\mciteSetBstMidEndSepPunct{\mcitedefaultmidpunct}
{\mcitedefaultendpunct}{\mcitedefaultseppunct}\relax
\EndOfBibitem
\bibitem[Dullien(1979)]{dullien}
Dullien,~F. A.~L. \emph{Porous Media, Fluid Transport and Pore structure};
  Academic: San Diego, 1979\relax
\mciteBstWouldAddEndPuncttrue
\mciteSetBstMidEndSepPunct{\mcitedefaultmidpunct}
{\mcitedefaultendpunct}{\mcitedefaultseppunct}\relax
\EndOfBibitem
\bibitem[Adler(1992)]{Adler}
Adler,~P.~M. \emph{Porous Media: Geometry and Transports};
  Butterworth-Heinemann: Stoneham, MA, USA, 1992\relax
\mciteBstWouldAddEndPuncttrue
\mciteSetBstMidEndSepPunct{\mcitedefaultmidpunct}
{\mcitedefaultendpunct}{\mcitedefaultseppunct}\relax
\EndOfBibitem
\bibitem[Guyont \latin{et~al.}(1987)Guyont, Oger, and Plona]{guyon1987}
Guyont,~E.; Oger,~L.; Plona,~T.~J. Transport properties in sintered porous
  media composed of two particle sizes. \emph{J. Phys. D: Appl. Phys.}
  \textbf{1987}, \emph{20}, 1637--1644\relax
\mciteBstWouldAddEndPuncttrue
\mciteSetBstMidEndSepPunct{\mcitedefaultmidpunct}
{\mcitedefaultendpunct}{\mcitedefaultseppunct}\relax
\EndOfBibitem
\bibitem[Kim and Torquato(1991)Kim, and Torquato]{kim1991}
Kim,~I.~C.; Torquato,~S. {Effective conductivity of suspensions of hard spheres
  by Brownian motion simulation}. \emph{J. Appl. Phys.} \textbf{1991},
  \emph{69}, 2280--2289\relax
\mciteBstWouldAddEndPuncttrue
\mciteSetBstMidEndSepPunct{\mcitedefaultmidpunct}
{\mcitedefaultendpunct}{\mcitedefaultseppunct}\relax
\EndOfBibitem
\bibitem[Coelho \latin{et~al.}(1997)Coelho, Thovert, and Adler]{coelho1997}
Coelho,~D.; Thovert,~J.-F.; Adler,~P.~M. Geometrical and transport properties
  of random packings of spheres and aspherical particles. \emph{Phys. Rev. E}
  \textbf{1997}, \emph{55}, 1959--1978\relax
\mciteBstWouldAddEndPuncttrue
\mciteSetBstMidEndSepPunct{\mcitedefaultmidpunct}
{\mcitedefaultendpunct}{\mcitedefaultseppunct}\relax
\EndOfBibitem
\bibitem[Tomadakis and Sotirchos(1993)Tomadakis, and Sotirchos]{tomadakis1993}
Tomadakis,~M.~M.; Sotirchos,~S. Ordinary and transition regime diffusion in
  random fiber structures. \emph{AIChE J.} \textbf{1993}, \emph{39},
  397--412\relax
\mciteBstWouldAddEndPuncttrue
\mciteSetBstMidEndSepPunct{\mcitedefaultmidpunct}
{\mcitedefaultendpunct}{\mcitedefaultseppunct}\relax
\EndOfBibitem
\bibitem[Sund \latin{et~al.}(2017)Sund, Porta, and Bolster]{sund2017}
Sund,~N.~L.; Porta,~G.~M.; Bolster,~D. Upscaling of dilution and mixing using a
  trajectory based Spatial Markov random walk model in a periodic flow domain.
  \emph{Adv. Water Res.} \textbf{2017}, \emph{103}, 76--85\relax
\mciteBstWouldAddEndPuncttrue
\mciteSetBstMidEndSepPunct{\mcitedefaultmidpunct}
{\mcitedefaultendpunct}{\mcitedefaultseppunct}\relax
\EndOfBibitem
\bibitem[{M\"uter} \latin{et~al.}(2015){M\"uter}, Sorensen, Bock, and
  Stipp]{muter2015}
{M\"uter},~D.; Sorensen,~H.~O.; Bock,~H.; Stipp,~S. L.~S. Particle Diffusion in
  Complex Nanoscale Pore Networks. \emph{J. Phys. Chem. C} \textbf{2015},
  \emph{119}, 10329--10335\relax
\mciteBstWouldAddEndPuncttrue
\mciteSetBstMidEndSepPunct{\mcitedefaultmidpunct}
{\mcitedefaultendpunct}{\mcitedefaultseppunct}\relax
\EndOfBibitem
\bibitem[Tallarek \latin{et~al.}(2019)Tallarek, Hlushkou, Rybka, and
  {H\"oltzel}]{tallarek2019}
Tallarek,~U.; Hlushkou,~D.; Rybka,~J.; {H\"oltzel},~A. Multiscale simulation of
  diffusion in porous media: {From} interfacial dynamics to hierarchical
  porosity. \emph{J. Phys. Chem. C} \textbf{2019}, \emph{123},
  15099--15112\relax
\mciteBstWouldAddEndPuncttrue
\mciteSetBstMidEndSepPunct{\mcitedefaultmidpunct}
{\mcitedefaultendpunct}{\mcitedefaultseppunct}\relax
\EndOfBibitem
\bibitem[Jiang \latin{et~al.}(2017)Jiang, Qu, Zhou, and Tao]{jiangApEn2017}
Jiang,~Z.~Y.; Qu,~Z.~G.; Zhou,~L.; Tao,~W.~Q. A microscopic investigation of
  ion and electron transport in lithium-ion battery porous electrodes using the
  lattice {Boltzmann} method. \emph{Applied Energy} \textbf{2017}, \emph{194},
  530--539\relax
\mciteBstWouldAddEndPuncttrue
\mciteSetBstMidEndSepPunct{\mcitedefaultmidpunct}
{\mcitedefaultendpunct}{\mcitedefaultseppunct}\relax
\EndOfBibitem
\bibitem[Torayev \latin{et~al.}(2018)Torayev, Magusin, Grey, Merlet, and
  Franco]{torayev2018}
Torayev,~A.; Magusin,~P. C. M.~M.; Grey,~C.~P.; Merlet,~C.; Franco,~A.~A.
  Importance of incorporating explicit {3D}-resolved electrode mesostructures
  in {Li-O${}_2$} battery models. \emph{ACS Appl. Energy Mater.} \textbf{2018},
  \emph{1}, 6433--6441\relax
\mciteBstWouldAddEndPuncttrue
\mciteSetBstMidEndSepPunct{\mcitedefaultmidpunct}
{\mcitedefaultendpunct}{\mcitedefaultseppunct}\relax
\EndOfBibitem
\bibitem[Elabyouki \latin{et~al.}(2019)Elabyouki, Bahamon, Khaleel, and
  Vega]{elabyouki2019}
Elabyouki,~M.; Bahamon,~D.; Khaleel,~M.; Vega,~L.~F. Insights into the
  transport properties of electrolyte solutions in a hierarchical carbon
  electrode by molecular dynamics simulations. \emph{J. Phys. Chem. C}
  \textbf{2019}, \emph{123}, 27273--27285\relax
\mciteBstWouldAddEndPuncttrue
\mciteSetBstMidEndSepPunct{\mcitedefaultmidpunct}
{\mcitedefaultendpunct}{\mcitedefaultseppunct}\relax
\EndOfBibitem
\bibitem[Tan \latin{et~al.}(2019)Tan, Kok, Daemi, Brett, and
  Shearing]{tanPCCP2019}
Tan,~C.; Kok,~M. D.~R.; Daemi,~S.~R.; Brett,~D. J.~L.; Shearing,~P.~R.
  Three-dimensional image based modelling of transport parameters in
  lithium-sulfur batteries. \emph{Phys. Chem. Chem. Phys.} \textbf{2019},
  \emph{21}, 4145--4154\relax
\mciteBstWouldAddEndPuncttrue
\mciteSetBstMidEndSepPunct{\mcitedefaultmidpunct}
{\mcitedefaultendpunct}{\mcitedefaultseppunct}\relax
\EndOfBibitem
\bibitem[Barab\'asi and Stanley(1995)Barab\'asi, and Stanley]{barabasi}
Barab\'asi,~A.-L.; Stanley,~H.~E. \emph{Fractal Concepts in Surface Growth};
  Cambridge University Press: New York, USA, 1995\relax
\mciteBstWouldAddEndPuncttrue
\mciteSetBstMidEndSepPunct{\mcitedefaultmidpunct}
{\mcitedefaultendpunct}{\mcitedefaultseppunct}\relax
\EndOfBibitem
\bibitem[Vold(1959)]{vold}
Vold,~M.~J. Sediment volume and structure in dispersions of anisometric
  particles. \emph{J. Phys. Chem.} \textbf{1959}, \emph{63}, 1608--1612\relax
\mciteBstWouldAddEndPuncttrue
\mciteSetBstMidEndSepPunct{\mcitedefaultmidpunct}
{\mcitedefaultendpunct}{\mcitedefaultseppunct}\relax
\EndOfBibitem
\bibitem[Family(1986)]{family}
Family,~F. Scaling of rough surfaces: {Effects} of surface diffusion. \emph{J.
  Phys. A: Math. Gen.} \textbf{1986}, \emph{19}, L441\relax
\mciteBstWouldAddEndPuncttrue
\mciteSetBstMidEndSepPunct{\mcitedefaultmidpunct}
{\mcitedefaultendpunct}{\mcitedefaultseppunct}\relax
\EndOfBibitem
\bibitem[Liu \latin{et~al.}(2017)Liu, {Wood III}, and Mukherjee]{liuPCCP2017}
Liu,~Z.; {Wood III},~D.~L.; Mukherjee,~P.~P. Evaporation induced nanoparticle -
  binder interaction in electrode film formation. \emph{Phys. Chem. Chem.
  Phys.} \textbf{2017}, \emph{19}, 10051--10061\relax
\mciteBstWouldAddEndPuncttrue
\mciteSetBstMidEndSepPunct{\mcitedefaultmidpunct}
{\mcitedefaultendpunct}{\mcitedefaultseppunct}\relax
\EndOfBibitem
\bibitem[Lehnen and Lu(2010)Lehnen, and Lu]{lehnen2010}
Lehnen,~C.; Lu,~T. Morphological evolution in ballistic deposition. \emph{Phys.
  Rev. E} \textbf{2010}, \emph{82}, 085437\relax
\mciteBstWouldAddEndPuncttrue
\mciteSetBstMidEndSepPunct{\mcitedefaultmidpunct}
{\mcitedefaultendpunct}{\mcitedefaultseppunct}\relax
\EndOfBibitem
\bibitem[Hoshen and Kopelman(1976)Hoshen, and Kopelman]{hoshen}
Hoshen,~J.; Kopelman,~R. Percolation and cluster distribution. {I. Cluster}
  multiple labeling technique and critical concentration algorithm. \emph{Phys.
  Rev. B} \textbf{1976}, \emph{14}, 3438--3445\relax
\mciteBstWouldAddEndPuncttrue
\mciteSetBstMidEndSepPunct{\mcitedefaultmidpunct}
{\mcitedefaultendpunct}{\mcitedefaultseppunct}\relax
\EndOfBibitem
\bibitem[Stauffer and Aharony(1992)Stauffer, and Aharony]{stauffer}
Stauffer,~D.; Aharony,~A. \emph{Introduction to Percolation Theory}; Taylor \&
  Francis: London/Philadelphia, 1992\relax
\mciteBstWouldAddEndPuncttrue
\mciteSetBstMidEndSepPunct{\mcitedefaultmidpunct}
{\mcitedefaultendpunct}{\mcitedefaultseppunct}\relax
\EndOfBibitem
\bibitem[{Aar\~ao Reis}(2016)]{reis2016}
{Aar\~ao Reis},~F. D.~A. Scaling relations in the diffusive infiltration in
  fractals. \emph{Phys. Rev. E} \textbf{2016}, \emph{94}, 052124\relax
\mciteBstWouldAddEndPuncttrue
\mciteSetBstMidEndSepPunct{\mcitedefaultmidpunct}
{\mcitedefaultendpunct}{\mcitedefaultseppunct}\relax
\EndOfBibitem
\bibitem[Reis and Voller(2019)Reis, and Voller]{reisvoller2019}
Reis,~F. D. A.~A.; Voller,~V.~R. Models of infiltration into homogeneous and
  fractal porous media with localized sources. \emph{Phys. Rev. E}
  \textbf{2019}, \emph{99}, 042111\relax
\mciteBstWouldAddEndPuncttrue
\mciteSetBstMidEndSepPunct{\mcitedefaultmidpunct}
{\mcitedefaultendpunct}{\mcitedefaultseppunct}\relax
\EndOfBibitem
\bibitem[Krug(1997)]{krug}
Krug,~J. Origins of scale invariance in growth processes. \emph{Adv. Phys.}
  \textbf{1997}, \emph{46}, 139--282\relax
\mciteBstWouldAddEndPuncttrue
\mciteSetBstMidEndSepPunct{\mcitedefaultmidpunct}
{\mcitedefaultendpunct}{\mcitedefaultseppunct}\relax
\EndOfBibitem
\bibitem[Kardar \latin{et~al.}(1986)Kardar, Parisi, and Zhang]{kpz}
Kardar,~M.; Parisi,~G.; Zhang,~Y.-C. Dynamic scaling of growing interfaces.
  \emph{Phys. Rev. Lett.} \textbf{1986}, \emph{56}, 889--892\relax
\mciteBstWouldAddEndPuncttrue
\mciteSetBstMidEndSepPunct{\mcitedefaultmidpunct}
{\mcitedefaultendpunct}{\mcitedefaultseppunct}\relax
\EndOfBibitem
\bibitem[Kelling \latin{et~al.}(2016)Kelling, \'Odor, and Gemming]{kelling2016}
Kelling,~J.; \'Odor,~G.; Gemming,~S. Universality of (2+1)-dimensional
  restricted solid-on-solid models. \emph{Phys. Rev. E} \textbf{2016},
  \emph{94}, 022107\relax
\mciteBstWouldAddEndPuncttrue
\mciteSetBstMidEndSepPunct{\mcitedefaultmidpunct}
{\mcitedefaultendpunct}{\mcitedefaultseppunct}\relax
\EndOfBibitem
\bibitem[{Kert\'esz} and Wolf(1988){Kert\'esz}, and Wolf]{kertesz}
{Kert\'esz},~J.; Wolf,~D.~E. Noise reduction in {Eden models: II. Surface
  structure and intrinsic width}. \emph{J. Phys. A: Math. Gen.} \textbf{1988},
  \emph{21}, 747--761\relax
\mciteBstWouldAddEndPuncttrue
\mciteSetBstMidEndSepPunct{\mcitedefaultmidpunct}
{\mcitedefaultendpunct}{\mcitedefaultseppunct}\relax
\EndOfBibitem
\bibitem[Alves \latin{et~al.}(2014)Alves, Oliveira, and Ferreira]{alvesPRE2014}
Alves,~S.~G.; Oliveira,~T.~J.; Ferreira,~S.~C. Origins of scaling corrections
  in ballistic growth models. \emph{Phys. Rev. E} \textbf{2014}, \emph{90},
  052405\relax
\mciteBstWouldAddEndPuncttrue
\mciteSetBstMidEndSepPunct{\mcitedefaultmidpunct}
{\mcitedefaultendpunct}{\mcitedefaultseppunct}\relax
\EndOfBibitem
\bibitem[Krug(1990)]{krug1990}
Krug,~J. Universal finite-size effects in the rate of growth processes.
  \emph{J. Phys. A: Math. Gen.} \textbf{1990}, \emph{23}, L987--L994\relax
\mciteBstWouldAddEndPuncttrue
\mciteSetBstMidEndSepPunct{\mcitedefaultmidpunct}
{\mcitedefaultendpunct}{\mcitedefaultseppunct}\relax
\EndOfBibitem
\bibitem[Edwards and Wilkinson(1982)Edwards, and Wilkinson]{ew}
Edwards,~S.~F.; Wilkinson,~D.~R. The surface statistics of a granular
  aggregate. \emph{Proc. R. Soc. Lond. A} \textbf{1982}, \emph{381},
  17--31\relax
\mciteBstWouldAddEndPuncttrue
\mciteSetBstMidEndSepPunct{\mcitedefaultmidpunct}
{\mcitedefaultendpunct}{\mcitedefaultseppunct}\relax
\EndOfBibitem
\bibitem[Pellegrini and Jullien(2000)Pellegrini, and Jullien]{pellegrini}
Pellegrini,~Y.~P.; Jullien,~R. Roughening transition and percolation in random
  ballistic deposition. \emph{Phys. Rev. Lett.} \textbf{2000}, \emph{64},
  1745--1748\relax
\mciteBstWouldAddEndPuncttrue
\mciteSetBstMidEndSepPunct{\mcitedefaultmidpunct}
{\mcitedefaultendpunct}{\mcitedefaultseppunct}\relax
\EndOfBibitem
\bibitem[Canet \latin{et~al.}(2010)Canet, Chat\'e, Delamotte, and
  Wschebor]{canetPRL2010}
Canet,~L.; Chat\'e,~H.; Delamotte,~B.; Wschebor,~N. Nonperturbative
  renormalization group for the {Kardar-Parisi-Zhang} equation. \emph{Phys.
  Rev. Lett.} \textbf{2010}, \emph{104}, 150601\relax
\mciteBstWouldAddEndPuncttrue
\mciteSetBstMidEndSepPunct{\mcitedefaultmidpunct}
{\mcitedefaultendpunct}{\mcitedefaultseppunct}\relax
\EndOfBibitem
\bibitem[Chame and Reis(2002)Chame, and Reis]{chamereis2002}
Chame,~A.; Reis,~F. D. A.~A. Crossover effects in a discrete deposition model
  with {Kardar-Parisi-Zhang} scaling. \emph{Phys. Rev. E} \textbf{2002},
  \emph{66}, 051104\relax
\mciteBstWouldAddEndPuncttrue
\mciteSetBstMidEndSepPunct{\mcitedefaultmidpunct}
{\mcitedefaultendpunct}{\mcitedefaultseppunct}\relax
\EndOfBibitem
\bibitem[Muraca \latin{et~al.}(2004)Muraca, Braunstein, and Buceta]{muraca2004}
Muraca,~D.; Braunstein,~L.~A.; Buceta,~R.~C. Universal behavior of the
  coefficients of the continuous equation in competitive growth models.
  \emph{Phys. Rev. E} \textbf{2004}, \emph{69}, 065103\relax
\mciteBstWouldAddEndPuncttrue
\mciteSetBstMidEndSepPunct{\mcitedefaultmidpunct}
{\mcitedefaultendpunct}{\mcitedefaultseppunct}\relax
\EndOfBibitem
\bibitem[Silveira and {Aar\~ao Reis}(2012)Silveira, and {Aar\~ao
  Reis}]{silveira2012}
Silveira,~F.~A.; {Aar\~ao Reis},~F. D.~A. Langevin equations for competitive
  growth models. \emph{Phys. Rev. E} \textbf{2012}, \emph{85}, 011601\relax
\mciteBstWouldAddEndPuncttrue
\mciteSetBstMidEndSepPunct{\mcitedefaultmidpunct}
{\mcitedefaultendpunct}{\mcitedefaultseppunct}\relax
\EndOfBibitem
\bibitem[Kriston \latin{et~al.}(2016)Kriston, Pfrang, and
  Boon-Brett]{kriston2016}
Kriston,~A.; Pfrang,~A.; Boon-Brett,~L. Development of multi-scale structure
  homogenization approaches based on modeled particle deposition for the
  simulation of electrochemical energy conversion and storage devices.
  \emph{Electrochim. Acta} \textbf{2016}, \emph{201}, 380--394\relax
\mciteBstWouldAddEndPuncttrue
\mciteSetBstMidEndSepPunct{\mcitedefaultmidpunct}
{\mcitedefaultendpunct}{\mcitedefaultseppunct}\relax
\EndOfBibitem
\bibitem[Katzav \latin{et~al.}(2006)Katzav, Edwards, and Schwartz]{katzav2006}
Katzav,~E.; Edwards,~S.~F.; Schwartz,~M. Structure below the growing surface.
  \emph{Europhys. Lett.} \textbf{2006}, \emph{75}, 29--35\relax
\mciteBstWouldAddEndPuncttrue
\mciteSetBstMidEndSepPunct{\mcitedefaultmidpunct}
{\mcitedefaultendpunct}{\mcitedefaultseppunct}\relax
\EndOfBibitem
\bibitem[Yu and Amar(2002)Yu, and Amar]{yuamar2002}
Yu,~J.; Amar,~J.~G. Scaling behavior of the surface in ballistic deposition.
  \emph{Phys. Rev. E} \textbf{2002}, \emph{65}, 060601(R)\relax
\mciteBstWouldAddEndPuncttrue
\mciteSetBstMidEndSepPunct{\mcitedefaultmidpunct}
{\mcitedefaultendpunct}{\mcitedefaultseppunct}\relax
\EndOfBibitem
\bibitem[Reis and di~Caprio(2014)Reis, and di~Caprio]{reiscaprio2014}
Reis,~F. D. A.~A.; di~Caprio,~D. Crossover from anomalous to normal diffusion
  in porous media. \emph{Phys. Rev. E} \textbf{2014}, \emph{89}, 062126\relax
\mciteBstWouldAddEndPuncttrue
\mciteSetBstMidEndSepPunct{\mcitedefaultmidpunct}
{\mcitedefaultendpunct}{\mcitedefaultseppunct}\relax
\EndOfBibitem
\bibitem[Cussler(2007)]{cussler}
Cussler,~E.~L. \emph{Diffusion: {Mass} Transfer in Fluid Systems}, 3rd ed.;
  Cambridge University Press: Cambridge, UK, 2007\relax
\mciteBstWouldAddEndPuncttrue
\mciteSetBstMidEndSepPunct{\mcitedefaultmidpunct}
{\mcitedefaultendpunct}{\mcitedefaultseppunct}\relax
\EndOfBibitem
\bibitem[Ben-Avraham and Havlin(2000)Ben-Avraham, and Havlin]{havlin}
Ben-Avraham,~D.; Havlin,~S. \emph{Diffusion and Reactions in Fractals and
  Disordered Systems}; Cambridge University Press: Cambridge, UK, 2000\relax
\mciteBstWouldAddEndPuncttrue
\mciteSetBstMidEndSepPunct{\mcitedefaultmidpunct}
{\mcitedefaultendpunct}{\mcitedefaultseppunct}\relax
\EndOfBibitem
\bibitem[Bouchaud and Georges(1990)Bouchaud, and Georges]{bouchaud}
Bouchaud,~J.~P.; Georges,~A. Anomalous diffusion in disordered media:
  Statistical mechanisms, models and physical applications. \emph{Phys. Rep.}
  \textbf{1990}, \emph{195}, 127--293\relax
\mciteBstWouldAddEndPuncttrue
\mciteSetBstMidEndSepPunct{\mcitedefaultmidpunct}
{\mcitedefaultendpunct}{\mcitedefaultseppunct}\relax
\EndOfBibitem
\bibitem[Metzler \latin{et~al.}(2014)Metzler, Jeon, Cherstvya, and
  Barkai]{metzler2014}
Metzler,~R.; Jeon,~J.-H.; Cherstvya,~A.~G.; Barkai,~E. Anomalous diffusion
  models and their properties: non-stationarity, non-ergodicity, and ageing at
  the centenary of single particle tracking. \emph{Phys. Chem. Chem. Phys.}
  \textbf{2014}, \emph{16}, 24128--24164\relax
\mciteBstWouldAddEndPuncttrue
\mciteSetBstMidEndSepPunct{\mcitedefaultmidpunct}
{\mcitedefaultendpunct}{\mcitedefaultseppunct}\relax
\EndOfBibitem
\bibitem[Grathwohl(1998)]{grathwohl}
Grathwohl,~P. \emph{Diffusion in natural porous media: contaminant transport,
  sorption/desorption and dissolution kinetics}; Springer: New York, USA,
  1998\relax
\mciteBstWouldAddEndPuncttrue
\mciteSetBstMidEndSepPunct{\mcitedefaultmidpunct}
{\mcitedefaultendpunct}{\mcitedefaultseppunct}\relax
\EndOfBibitem
\bibitem[Archie(1942)]{archie1942}
Archie,~G. : The electrical resistivity log as an aid in determining some
  reservoir characteristics. \emph{Trans. AIME} \textbf{1942}, \emph{146},
  54--61\relax
\mciteBstWouldAddEndPuncttrue
\mciteSetBstMidEndSepPunct{\mcitedefaultmidpunct}
{\mcitedefaultendpunct}{\mcitedefaultseppunct}\relax
\EndOfBibitem
\bibitem[Havlin \latin{et~al.}(1983)Havlin, ben Avraham, and
  Sompolinsky]{havlin1983b}
Havlin,~S.; ben Avraham,~D.; Sompolinsky,~H. Scaling behavior of diffusion on
  percolation clusters. \emph{Phys. Rev. A} \textbf{1983}, \emph{27},
  1730--1733\relax
\mciteBstWouldAddEndPuncttrue
\mciteSetBstMidEndSepPunct{\mcitedefaultmidpunct}
{\mcitedefaultendpunct}{\mcitedefaultseppunct}\relax
\EndOfBibitem
\bibitem[Gao \latin{et~al.}(2018)Gao, Wu, Hu, Zheng, Amine, and Chen]{gao2018}
Gao,~H.; Wu,~Q.; Hu,~Y.; Zheng,~J.~P.; Amine,~K.; Chen,~Z. Revealing the
  rate-limiting {Li-ion} diffusion pathway in ultrathick electrodes for
  {Li-ion} batteries. \emph{J. Phys. Chem. Lett.} \textbf{2018}, \emph{9},
  5100--5104\relax
\mciteBstWouldAddEndPuncttrue
\mciteSetBstMidEndSepPunct{\mcitedefaultmidpunct}
{\mcitedefaultendpunct}{\mcitedefaultseppunct}\relax
\EndOfBibitem
\bibitem[Hossain \latin{et~al.}(2019)Hossain, Stephens, Hatami, Ghavidel,
  Chhin, Dawkins, Savignac, Mauzeroll, and Schougaard]{hossain2019}
Hossain,~M.~S.; Stephens,~L.~I.; Hatami,~M.; Ghavidel,~M.; Chhin,~D.;
  Dawkins,~J. I.~G.; Savignac,~L.; Mauzeroll,~J.; Schougaard,~S.~B. Effective
  mass transport properties in lithium battery electrodes. \emph{ACS Appl.
  Energy Mater.} \textbf{2019}, \emph{3}, 440--446\relax
\mciteBstWouldAddEndPuncttrue
\mciteSetBstMidEndSepPunct{\mcitedefaultmidpunct}
{\mcitedefaultendpunct}{\mcitedefaultseppunct}\relax
\EndOfBibitem
\bibitem[Dai and Srinivasan(2016)Dai, and Srinivasan]{dai2016}
Dai,~Y.; Srinivasan,~V. On graded electrode porosity as a design tool for
  improving the energy density of batteries. \emph{J. Electrochem. Soc.}
  \textbf{2016}, \emph{163}, A406--A416\relax
\mciteBstWouldAddEndPuncttrue
\mciteSetBstMidEndSepPunct{\mcitedefaultmidpunct}
{\mcitedefaultendpunct}{\mcitedefaultseppunct}\relax
\EndOfBibitem
\bibitem[Colclasure \latin{et~al.}(2020)Colclasure, Tanim, Jansen, Trask,
  Dunlop, Polzin, Bloom, Robertson, Flores, Evans, Dufek, and
  Smith]{colclasure2020}
Colclasure,~A.~M.; Tanim,~T.~R.; Jansen,~A.~N.; Trask,~S.~E.; Dunlop,~A.~R.;
  Polzin,~B.~J.; Bloom,~I.; Robertson,~D.; Flores,~L.; Evans,~M. \latin{et~al.}
   Electrode scale and electrolyte transport effects on extreme fast charging
  of lithium-ion cells. \emph{Electrochim. Acta} \textbf{2020}, \emph{337},
  135854\relax
\mciteBstWouldAddEndPuncttrue
\mciteSetBstMidEndSepPunct{\mcitedefaultmidpunct}
{\mcitedefaultendpunct}{\mcitedefaultseppunct}\relax
\EndOfBibitem
\bibitem[Su \latin{et~al.}(2020)Su, {De Andrade}, Cretu, Yin, Wojcik, Franco,
  and {Demorti\`ere}]{suACSAMI2020}
Su,~Z.; {De Andrade},~V.; Cretu,~S.; Yin,~Y.; Wojcik,~M.~J.; Franco,~A.~A.;
  {Demorti\`ere},~A. X-ray nanocomputed tomography in {Zernike} phase contrast
  for studying {3D} morphology of {Li-O${}_2$} battery electrode. \emph{ACS
  Appl. Energy Mater.} \textbf{2020}, \emph{3}, 4093--4102\relax
\mciteBstWouldAddEndPuncttrue
\mciteSetBstMidEndSepPunct{\mcitedefaultmidpunct}
{\mcitedefaultendpunct}{\mcitedefaultseppunct}\relax
\EndOfBibitem
\bibitem[Vierrath \latin{et~al.}(2015)Vierrath, Zielke, Moroni, Mondon,
  Wheeler, Zengerle, and Thiele]{vierrath2015}
Vierrath,~S.; Zielke,~L.; Moroni,~R.; Mondon,~A.; Wheeler,~D.~R.; Zengerle,~R.;
  Thiele,~S. Morphology of nanoporous carbon binder domains in {Li-ion
  batteries - A FIB-SEM} study. \emph{Electrochem. Comm.} \textbf{2015},
  \emph{60}, 176--179\relax
\mciteBstWouldAddEndPuncttrue
\mciteSetBstMidEndSepPunct{\mcitedefaultmidpunct}
{\mcitedefaultendpunct}{\mcitedefaultseppunct}\relax
\EndOfBibitem
\bibitem[Usseglio-Viretta \latin{et~al.}(2018)Usseglio-Viretta, Colclasure,
  Mistry, Claver, Pouraghajan, Finegan, Heenan, Abraham, Mukherjee, Wheeler,
  Shearing, Cooper, and Smith]{viretta2018}
Usseglio-Viretta,~F. L.~E.; Colclasure,~A.; Mistry,~A.~N.; Claver,~K. P.~Y.;
  Pouraghajan,~F.; Finegan,~D.~P.; Heenan,~T. M.~M.; Abraham,~D.;
  Mukherjee,~P.~P.; Wheeler,~D. \latin{et~al.}  Resolving the discrepancy in
  tortuosity factor estimation for {Li}-ion battery electrodes through
  micro-macro modeling and experiment. \emph{J. Electrochem. Soc.}
  \textbf{2018}, \emph{165}, A3403--A3426\relax
\mciteBstWouldAddEndPuncttrue
\mciteSetBstMidEndSepPunct{\mcitedefaultmidpunct}
{\mcitedefaultendpunct}{\mcitedefaultseppunct}\relax
\EndOfBibitem
\bibitem[Liu \latin{et~al.}(2017)Liu, Guan, and Liu]{liuJES2017}
Liu,~L.; Guan,~P.; Liu,~C. Experimental and simulation investigations of
  porosity graded cathodes in mitigating battery degradation of high voltage
  lithium-ion batteries. \emph{J. Electrochem. Soc.} \textbf{2017}, \emph{164},
  A3163--A3173\relax
\mciteBstWouldAddEndPuncttrue
\mciteSetBstMidEndSepPunct{\mcitedefaultmidpunct}
{\mcitedefaultendpunct}{\mcitedefaultseppunct}\relax
\EndOfBibitem
\bibitem[Cheng \latin{et~al.}(2020)Cheng, Drummond, Duncan, and
  Grant]{cheng2020}
Cheng,~C.; Drummond,~R.; Duncan,~S.~R.; Grant,~P.~S. Combining composition
  graded positive and negative electrodes for higher performance {Li-ion}
  batteries. \emph{J. Power Sources} \textbf{2020}, \emph{448}, 227376\relax
\mciteBstWouldAddEndPuncttrue
\mciteSetBstMidEndSepPunct{\mcitedefaultmidpunct}
{\mcitedefaultendpunct}{\mcitedefaultseppunct}\relax
\EndOfBibitem
\bibitem[Zhan \latin{et~al.}(2020)Zhan, Edison, Manalastas, Tan, Buffa,
  Madhavi, and Mandler]{zhan2020}
Zhan,~Y.; Edison,~E.; Manalastas,~W.; Tan,~M. R.~J.; Buffa,~R. S.~A.;
  Madhavi,~S.; Mandler,~D. Electrochemical deposition of highly porous reduced
  graphene oxide electrodes for {Li-ion} capacitors. \emph{Electrochim. Acta}
  \textbf{2020}, \emph{337}, 135861\relax
\mciteBstWouldAddEndPuncttrue
\mciteSetBstMidEndSepPunct{\mcitedefaultmidpunct}
{\mcitedefaultendpunct}{\mcitedefaultseppunct}\relax
\EndOfBibitem
\bibitem[Das~Sarma and Tamborenea(1991)Das~Sarma, and Tamborenea]{dt}
Das~Sarma,~S.; Tamborenea,~P. A new universality class for kinetic growth:
  One-dimensional molecular-beam epitaxy. \emph{Phys. Rev. Lett.}
  \textbf{1991}, \emph{66}, 325--328\relax
\mciteBstWouldAddEndPuncttrue
\mciteSetBstMidEndSepPunct{\mcitedefaultmidpunct}
{\mcitedefaultendpunct}{\mcitedefaultseppunct}\relax
\EndOfBibitem
\bibitem[Wolf and Villain(1990)Wolf, and Villain]{wv}
Wolf,~D.~E.; Villain,~J. Growth with Surface Diffusion. \emph{EPL (Europhysics
  Letters)} \textbf{1990}, \emph{13}, 389\relax
\mciteBstWouldAddEndPuncttrue
\mciteSetBstMidEndSepPunct{\mcitedefaultmidpunct}
{\mcitedefaultendpunct}{\mcitedefaultseppunct}\relax
\EndOfBibitem
\bibitem[Chame and Reis(2004)Chame, and Reis]{chamereis2004}
Chame,~A.; Reis,~F. D. A.~A. Scaling of local interface width of statistical
  growth models. \emph{Surface Science} \textbf{2004}, \emph{553}, 145 --
  154\relax
\mciteBstWouldAddEndPuncttrue
\mciteSetBstMidEndSepPunct{\mcitedefaultmidpunct}
{\mcitedefaultendpunct}{\mcitedefaultseppunct}\relax
\EndOfBibitem
\bibitem[Aryanfar \latin{et~al.}(2015)Aryanfar, Cheng, Colussi, Merinov,
  {Goddard III}, and Hoffmann]{aryanfarJCP2015}
Aryanfar,~A.; Cheng,~T.; Colussi,~A.~J.; Merinov,~B.~V.; {Goddard III},~W.~A.;
  Hoffmann,~M.~R. Annealing kinetics of electrodeposited lithium dendrites.
  \emph{J. Chem. Phys.} \textbf{2015}, \emph{143}, 134701\relax
\mciteBstWouldAddEndPuncttrue
\mciteSetBstMidEndSepPunct{\mcitedefaultmidpunct}
{\mcitedefaultendpunct}{\mcitedefaultseppunct}\relax
\EndOfBibitem
\bibitem[Reis \latin{et~al.}(2017)Reis, di~Caprio, and
  Taleb]{reiscapriotaleb2017}
Reis,~F. D. A.~A.; di~Caprio,~D.; Taleb,~A. Crossover from compact to branched
  films in electrodeposition with surface diffusion. \emph{Phys. Rev. E}
  \textbf{2017}, \emph{96}, 022805\relax
\mciteBstWouldAddEndPuncttrue
\mciteSetBstMidEndSepPunct{\mcitedefaultmidpunct}
{\mcitedefaultendpunct}{\mcitedefaultseppunct}\relax
\EndOfBibitem
\bibitem[Vishnugopi \latin{et~al.}(2020)Vishnugopi, Hao, Verma, and
  Mukherjee]{vishnugopi2020}
Vishnugopi,~B.~S.; Hao,~F.; Verma,~A.; Mukherjee,~P.~P. Surface diffusion
  manifestation in electrodeposition of metal anodes. \emph{Phys. Chem. Chem.
  Phys.} \textbf{2020}, \emph{22}, 11286--11295\relax
\mciteBstWouldAddEndPuncttrue
\mciteSetBstMidEndSepPunct{\mcitedefaultmidpunct}
{\mcitedefaultendpunct}{\mcitedefaultseppunct}\relax
\EndOfBibitem
\bibitem[Giri \latin{et~al.}(2013)Giri, Tarafdar, Gouze, and Dutta]{giri2013}
Giri,~A.; Tarafdar,~S.; Gouze,~P.; Dutta,~T. Fractal geometry of sedimentary
  rocks: simulation in {3-D} using a Relaxed Bidisperse Ballistic Deposition
  Model. \emph{Geophys. J. Int.} \textbf{2013}, \emph{192}, 1059--1069\relax
\mciteBstWouldAddEndPuncttrue
\mciteSetBstMidEndSepPunct{\mcitedefaultmidpunct}
{\mcitedefaultendpunct}{\mcitedefaultseppunct}\relax
\EndOfBibitem
\end{mcitethebibliography}

\begin{thebibliography}{1}

\bibitem{hoshen}
J.~Hoshen and R.~Kopelman.
\newblock Percolation and cluster distribution. {I. Cluster} multiple labeling
  technique and critical concentration algorithm.
\newblock {\em Phys. Rev. B}, 14:3438--3445, 1976.

\bibitem{yuamar2002}
Jianguo Yu and Jacques~G. Amar.
\newblock Scaling behavior of the surface in ballistic deposition.
\newblock {\em Phys. Rev. E}, 65:060601(R), 2002.

\bibitem{havlin1983b}
S.~Havlin, D.~ben Avraham, and H.~Sompolinsky.
\newblock Scaling behavior of diffusion on percolation clusters.
\newblock {\em Phys. Rev. A}, 27:1730--1733, 1983.

\bibitem{havlinAdvPhys2002}
S.~Havlin and D.~{Ben-Avraham}.
\newblock Diffusion in disordered media*.
\newblock {\em Adv. Phys.}, 51:187--292, 2002.

\bibitem{stauffer}
D.~Stauffer and A.~Aharony.
\newblock {\em Introduction to Percolation Theory}.
\newblock Taylor \& Francis, London/Philadelphia, 1992.

\bibitem{normand1995}
J.~M. Normand and H.~J. Herrmann.
\newblock Precise determination of the conductivity exponent of {3D}
  percolation using {"Percola"}.
\newblock {\em Int. J. Mod. Phys. C}, 6:813--817, 1995.

\bibitem{jan1998}
N.~Jan and D.~Stauffer.
\newblock Random site percolation in three dimensions.
\newblock {\em Int. J. Mod. Phys. C}, 9:341--347, 1998.

\bibitem{grathwohl}
P.~Grathwohl.
\newblock {\em Diffusion in natural porous media: contaminant transport,
  sorption/desorption and dissolution kinetics}.
\newblock Springer, New York, USA, 1998.

\end{thebibliography}

\end{document}